\documentclass[aps, prl, preprint, graphics, showpacs, amssymb, amsmath, superscriptaddress]{revtex4-1}

\usepackage{xcolor}
\usepackage{soul}
\usepackage{graphicx}
\usepackage{hyperref}
\usepackage{float}          
\usepackage[version=4]{mhchem}
\usepackage{booktabs}
\usepackage{multirow}

\begin{document}
	
	\title{Computational Discovery of Metastable NaMnO$_2$ Polymorphs as High-Performance Cathodes with Ultralow Na$^+$ Migration Barriers}
	
	\author{Fukuan Wang}\affiliation{State Key Laboratory of Metastable Materials Science and Technology \& Hebei Key Laboratory of Microstructural Material Physics, School of Science, Yanshan University, Qinhuangdao 066004, China}
	\author{Chen Zhou}\affiliation{State Key Laboratory of Metastable Materials Science and Technology \& Hebei Key Laboratory of Microstructural Material Physics, School of Science, Yanshan University, Qinhuangdao 066004, China}
	\author{Busheng Wang}\email{bushengw@ysu.edu.cn}\affiliation{State Key Laboratory of Metastable Materials Science and Technology \& Hebei Key Laboratory of Microstructural Material Physics, School of Science, Yanshan University, Qinhuangdao 066004, China}
	\author{Yong Liu}\email{yongliu@ysu.edu.cn}\affiliation{State Key Laboratory of Metastable Materials Science and Technology \& Hebei Key Laboratory of Microstructural Material Physics, School of Science, Yanshan University, Qinhuangdao 066004, China}
	
	\date{\today}
	
	\begin{abstract}
		Using an \textit{ab initio} evolutionary algorithm combined with first-principles calculations, two metastable NaMnO\textsubscript{2} polymorphs, $I4_1/amd$ and \textit{Cmcm}, are identified as promising cathode materials for sodium-ion batteries. Both phases exhibit excellent thermodynamic stability, lying within 35~meV/atom of the ground-state \textit{Pmmn} phase across 0--50~GPa, and are dynamically and thermally stable under ambient conditions following high-pressure synthesis, as confirmed by phonon and \textit{ab initio} molecular dynamics simulations. During desodiation, a Jahn--Teller-induced magnetic transition enhances Mn--O hybridization, reduces the bandgap, and promotes robust charge compensation and oxygen retention. Remarkably, the \textit{Cmcm} phase achieves record-low Na$^+$ migration barriers (0.39~eV at high Na concentration; 0.27~eV at low concentration), representing 47\% and 36\% reductions respectively compared to conventional $C2/m$, while delivering a higher average voltage (3.19~V vs 2.88~V). The $I4_1/amd$ phase exhibits concentration-dependent diffusion with a low-energy pathway (0.38~eV) and maintains competitive voltage (2.94~V). These findings suggest that metastable NaMnO$_2$ polymorphs may offer viable alternatives to conventional cathode materials, particularly where fast ionic conduction is required.
	\end{abstract}
	
	\maketitle
	
	\section{INTRODUCTION}
	
	With the growing demand for energy storage in portable electronics and grid-scale applications, lithium-ion batteries (LIBs) have dominated the market due to their high energy density and conversion efficiency. However, limited lithium resources, increasing costs, and environmental concerns hinder their long-term sustainability\cite{Li-limited,Li-limited-2020,Li-limited-2022,Li-limited-2024}. In contrast, sodium-ion batteries (SIBs) have garnered significant attention as promising alternatives, benefiting from sodium's abundance, low cost, and environmental friendliness\cite{slater2013sodium,yabuuchi2014research,hwang2017sodium,nayak2018lithium,vaalma2018cost,zhao2023recycling}. A wide range of cathode materials have been proposed for SIBs, including layered transition metal oxides\cite{NaxTMO2,toumar2015vacancy,wang2019ni,liu2023challenges}, polyanionic compounds\cite{juyinlizi}, Prussian blue analogs\cite{prussianblue}, tunnel oxides\cite{liu2024reviving,xiao2025guideline}, and organic molecules/polymers\cite{zeng2025crystalline}.
	
	Layered sodium manganese oxides (Na$_x$MnO$_2$) are among the most promising cathode candidates for SIBs, owing to the abundance and low cost of manganese, alongside their high theoretical capacities and favorable safety profiles\cite{Mn-NaMnO2,bi2021rechargeable,zhang2025crystal}. These compounds exhibit characteristic P2, P$^\prime$2, and O$^\prime$3 polymorphs, where the nomenclature specifies Na$^+$ coordination environment (prismatic or octahedral) and the stacking sequence of transition metal oxide layers\cite{name}. The prime notation indicates structural distortions induced by Jahn-Teller (JT) active Mn$^{3+}$ ions, which profoundly influence the electrochemical behavior.
	
	Among these polymorphs, P-type phases, such as P$^\prime$2-Na$_{0.67}$MnO$_2$\cite{P'2}, are particularly attractive due to their open prismatic pathways for Na$^+$ diffusion, which facilitate enhanced rate capability and reversible capacity.\cite{Cmcm-DFT-pccp-2020} However, their relatively low sodium content ($x < 1$) limits achievable energy density and complicates full-cell integration\cite{PvsO}. In contrast, fully sodiated NaMnO$_2$ ($x=1$) phases---including the monoclinic O$^\prime$3-type and orthorhombic zigzag-type---offer higher sodium inventory and capacity but suffer from sluggish ionic diffusion, as Na$^+$ must traverse energetically unfavorable tetrahedral sites during migration\cite{O'3,beita,NaxTMO2}. These phases are also sensitive to air and exhibit poorer long-term cyclability.
	
	Efforts to resolve these challenges have typically focused on chemical strategies, such as elemental doping and substitution, which aim to regulate the Mn$^{3+}$/Mn$^{4+}$ ratio, suppress JT distortions, enhance structural robustness, and engineer favorable diffusion pathways\cite{npj-2024}. While effective to some extent, these approaches introduce additional compositional complexity and cost, and often result in metastable local structures whose formation and persistence under ambient conditions remain difficult to control or predict\cite{chanza-de-quedian}. This underscores the need for alternative, physically grounded strategies for phase stabilization and performance enhancement.
	
	In this context, high-pressure synthesis has recently gained attention as a viable route to access thermodynamically inaccessible or metastable phases with improved electrochemical and structural properties\cite{Matter-2021}. For example, Liu \textit{et al.} demonstrated that applying pressure can suppress JT distortions in layered Mn oxides, leading to enhanced symmetry and cycling stability. Uyama \textit{et al.} synthesized a high-pressure $\gamma$-LiFeO$_2$-type tetragonal LiMnO$_2$ above 8 GPa, which displayed improved energy density relative to conventional LiMnO$_2$\cite{LiMnO2}. High pressure can tune local coordination environments and Mn--O bond lengths, offering a promising tool for crystal engineering of transition metal oxides.
	
	Nevertheless, the experimental realization of high-pressure phases remains time- and resource-intensive, often requiring specialized equipment and extensive synthesis-optimization cycles. To circumvent these bottlenecks, the integration of \textit{ab initio} crystal structure prediction methods have proven powerful in accelerating the discovery of novel battery materials\cite{wangshuo-Li2MnSiO4,wangshuo-LiCoO2,wangshuo-Li2MnO3}.  For example, Wang \textit{et al.} predicted high-pressure phases of Li$_2$MnSiO$_4$ with fast ion transport\cite{wangshuo-Li2MnSiO4}, rutile-like Li$_x$CoO$_2$ phases with superior stability\cite{wangshuo-LiCoO2}, and a non-layered metastable Li$_2$MnO$_3$ phase with enhanced electronic response and energy density\cite{wangshuo-Li2MnO3}. While such efforts have significantly advanced the design of lithium-based cathode materials, sodium-based analogues remain underexplored, particularly for Mn-based systems that exhibit rich polymorphism and complex JT physics under pressure.
	
	In this work, we employ a global \textit{ab initio} evolutionary algorithm (EA) and density functional theory (DFT) calculations to explore high-pressure-induced polymorphs of NaMnO$_2$ as next-generation cathode materials for SIBs. Two metastable structures, rocksalt-type $I4_1/amd$ and orthorhombic \textit{Cmcm}, are identified within 35 meV/atom of the ground-state \textit{Pmmn} phase across a wide pressure range (0--50 GPa). Both structures are dynamically and thermally stable under ambient conditions, as confirmed by phonon dispersion and \textit{ab initio} molecular dynamics simulations. We systematically evaluate their potential for battery applications, including electronic structure evolution during desodiation, magnetic phase transitions, Mn--O hybridization-driven charge compensation, Na$^+$ migration barriers at varying concentrations, open-circuit voltages (OCV), and oxygen vacancy formation energies. Our results underscore the synergistic advantages of these two metastable NaMnO$_2$ phases, including fast sodium-ion diffusion, stable voltage profiles, and excellent cycling durability, positioning them as strong candidates for high-performance SIB cathodes.
	
	\section{METHODS}
	
	Crystal structure prediction searches were carried out using the  XTALOPT \cite{lonie2011xtalopt} evolutionary algorithm release 13 \cite{hajinazar2024xtalopt}, which was designed to find stable and metastable structures given only their composition. Systematic searches were performed for systems containing up to 16 atoms per primitive cell (considering $Z$ = 1, 2, 3, and 4) at pressures of 0, 10, 30, and 50 GPa. Subsequent geometry optimizations and electronic structure calculations were carried out using the Vienna \textit{ab initio} simulation package (VASP)~\cite{VASP,VASP2}, employing the Perdew-Burke-Ernzerhof (PBE) generalized gradient approximation~\cite{PBE1,PBE2}. The electron-ion interactions were treated within the projector-augmented-wave (PAW) framework~\cite{PAW}, with valence electron configurations of Na ($3s^1$), Mn ($3d^6$$4s^1$), and O ($2s^2$$2p^4$). We used a plane-wave energy cutoff of 520 eV and a Monkhorst-Pack $k$-point grid~\cite{KPOINTS} with a spacing of 0.03 \AA$^{-1}$. To accurately describe the strong electron correlations within the Mn $3d$ orbitals, we adopted the GGA+$U$ approach with $U_{\text{eff}}$ = 3.9 eV for Mn$^{3+}$ ions, including spin polarization effects~\cite{U1-cengzhuangnengleicankao,U2-compare,U3}. To determine the magnetic ground states of phases, we systematically  evaluated the influence of magnetic ordering---including nonmagnetic (NM), ferromagnetic (FM), and several antiferromagnetic (AFM) configurations---on phase stability~\cite{AFMcankao1,Ceder-2025-LiMnO2}. 
	
	Phonon dispersions were computed using the finite displacement method as implemented in the phonopy code~\cite{Phonopy}. Thermal stability was evaluated through \textit{ab initio} molecular dynamics (AIMD) simulations in the canonical (NVT) ensemble at 500 K, employing a Nos\'e-Hoover thermostat~\cite{Nose-Hoover} with a 1 fs time step for 10 ps. 
	To accurately account for dispersion effects, we employed the vdW-D3 correction scheme~\cite{VDW-D3}, which has proven reliable for capturing van der Waals interactions in layered oxides ~\cite{Cmcm-DFT-pccp-2020,npj-2024}. Sodium ion migration barriers were calculated using the climbing-image nudged elastic band (CI-NEB) method~\cite{CI-NEB} in a 128-atom supercell (32 formula units). Following established methodology~\cite{npj-ceder-2016}, fixed atomic charges were assumed to enumerate 10 symmetrically inequivalent configurations with the lowest electrostatic energies using supercell program~\cite{supercell-program}. These configurations were subsequently optimized through first-principles calculations, with the lowest-energy structure selected as the ground state for each  varying Na concentrations. The enumeration process employed the same 128-atom supercell (32 formula units).

	\section{RESULTS}
	\subsection{Structure and stability}
	We conducted a global structural search for conducted across 0-50 GPa, considering NM, FM and AFM configurations in the post-processing. Four candidate phases with AFM ground states were identified, their crystal structures and relative enthalpy difference are illustrated in Figure~\ref{FIG-1}, with detailed configurations and lattice parameters provided in the Supplemental Material (Table S2). Experimentally, NaMnO\(_2\) predominantly crystallizes in the \textit{Pmmn} (zigzag-type \(\beta\)-phase) and $C2/m$ (layered \(\alpha\)-phase) structures, as shown in Figure~\ref{FIG-1}(a-b). Our calculations confirm the \textit{Pmmn} phase as the ground state, with a energy difference of $\sim$2 meV/atom relative to $C2/m$---consistent with observed phase mixing\cite{AFMcankao3-compare-daixicankao,beita,U2-compare}. These results validate the EA search for the NaMnO$_2$ system under pressure.
	
	\begin{figure}[hbtp]
		\centering
		\rotatebox{0}{\includegraphics[width=0.65\textwidth]{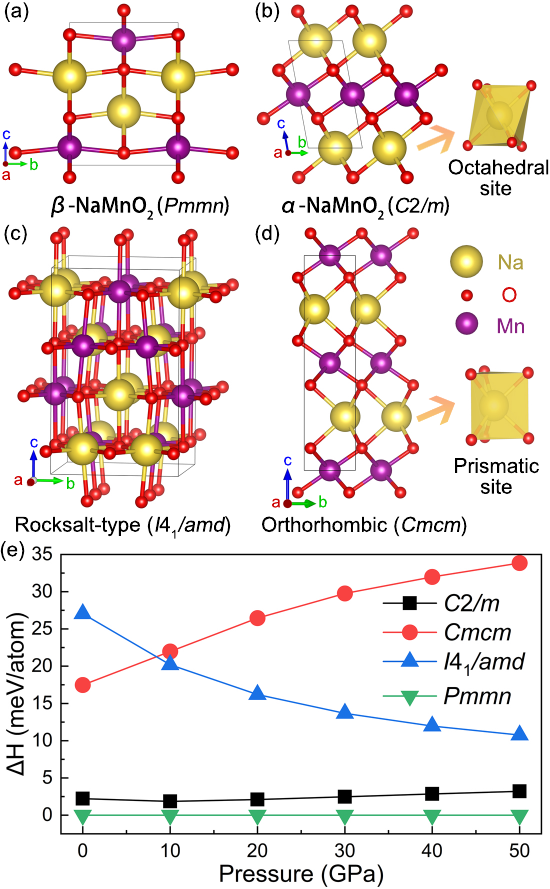}}
		\caption{(a-d) Crystal structures of NaMnO$_2$ polymorphs: (a) Zigzag-type $\beta$-NaMnO$_2$ (\textit{Pmmn}), (b) layered $\alpha$-NaMnO$_2$ ($C2/m$), (c) Rocksalt-type $I4_1/amd$, and (d) orthorhombic \textit{Cmcm}. Yellow, red and purple spheres represent Na, O and Mn atoms, respectively. Distinct NaO$_6$ coordination modes in (b) and (d) are highlighted with yellow arrows. (e) Enthalpy difference ($\Delta H$) versus pressure showing $\beta$-NaMnO$_2$ \textit{Pmmn} as the ground state (green circles), with a $\sim$2 meV/atom difference from $C2/m$--consistent with observed phase mixing~\cite{U2-compare,AFMcankao3-compare-daixicankao}, while $I4_1/amd$ and \textit{Cmcm} remain metastable. The $I4_1/amd$ phase becomes competitive ($\Delta H$ \textless 15 meV/atom) above 20 GPa.}\label{FIG-1}
	\end{figure}
	
	Beyond the experimentally characterized $\beta$-NaMnO$_2$ (\textit{Pmmn}) and $\alpha$-NaMnO$_2$ ($C2/m$) phases, we identify two metastable configurations: (i) a Rocksalt-type $I4_1/amd$ structure featuring $\lambda$-MnO$_2$-type Mn frameworks with cubic close-packed (ccp) oxygen sublattices and NaO$_6$ octahedral coordination, and (ii) an orthorhombic \textit{Cmcm} phase (Figure~\ref{FIG-1}c, d). Pressure-dependent analysis reveals the $I4_1/amd$ phase approaches energetic degeneracy ($\Delta H$ = 11 meV/atom at 50 GPa) with the \textit{Pmmn} ground state (Fig.~\ref{FIG-1}e), attributable to pressure-induced suppression of JT distortions in MnO$_6$ octahedra. As pressure increases, the resulting compression simultaneously reduces Mn-O bond length asymmetry while disrupting the long-range correlation of JT distortions. Consequently, the system's cooperative distortion becomes inhibited, leading to significant enthalpy reduction. The narrow 11 meV/atom energy difference at high pressure suggests the tetragonal phase could be stabilized through shock-wave or dynamic compression techniques\cite{High-pressure-shiyan}. 
	
	\begin{figure}[hbtp]
	\centering
	\includegraphics[height=16cm]{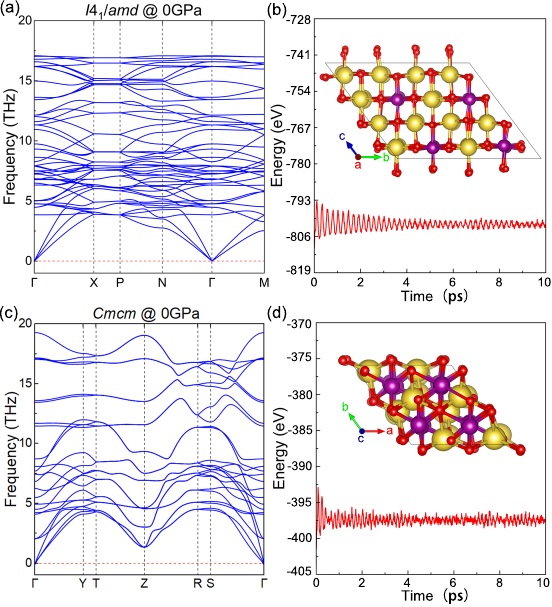}
	\caption{Phonon dispersion for (a) $I4_1/amd$ and (c) \textit{Cmcm} NaMnO$_2$ at ambient pressure. The absence of imaginary frequencies confirms their dynamic stability. AIMD simulations at 500 K demonstrate thermal stability of (b) $I4_1/amd$ and (d) \textit{Cmcm} phases through total energy fluctuations over 10 ps. Insets show the preserved coordination in final configurations after 10 ps simulations. \label{FIG-2}}
	\end{figure}
	
	The \textit{Cmcm} phase of $\mathrm{Na_{x}MnO_{2}}$ ($x \approx 2/3$) has been experimentally realized \cite{Cmcm-DFT-pccp-2020}, exhibiting trigonal prismatic $\mathrm{NaO_{6}}$ coordination. While experimental studies report both face-sharing and edge-sharing configurations between $\mathrm{NaO_{6}}$ prisms and $\mathrm{MnO_{6}}$ octahedra, our calculations show exclusive edge-sharing connectivity due to electrostatic repulsion minimization between $\mathrm{Na^{+}}$ and $\mathrm{Mn^{3+}}$ ions in face-sharing arrangements. Remarkably, this phase demonstrates near-degeneracy with the \textit{Pmmn} ground state at ambient pressure ($\Delta H = 17$ meV/atom). Under high pressure, increased O-O repulsion from interlayer compression raises the energy difference to 34 meV/atom at 50 GPa. This modest energy penalty suggests viable low-pressure synthesis pathways. The persistence of edge-sharing connectivity across the pressure range reveals fundamental geometric constraints in prismatic coordination. These results demonstrate how competing energetics (electrostatic repulsion vs. geometric strain) govern the structural evolution of NaMnO$_2$ polymorphs, providing design principles for targeted synthesis of metastable phases through pressure modulation.
	
	To systematically investigate the stability of the metastable $I4_1/amd$ and \textit{Cmcm} polymorphs, we performed complementary first-principles analyses combining phonon calculations and AIMD simulations. The phonon dispersion spectra for $I4_1/amd$ at 30~GPa and \textit{Cmcm} at 50~GPa show no imaginary frequencies (Fig.S5 in Supplemental Material), confirming their dynamic stability under pressure. Remarkably, this stability persists even at ambient conditions (Fig.~\ref{FIG-1} a, c), suggesting these phases could potentially be recovered through pressure quenching techniques. To further assess thermal stability, we conducted AIMD simulations at 500~K using $2\times2\times2$ supercells with 10~ps trajectories. The results demonstrate excellent thermal stability, as evidenced by only minimal total energy fluctuations (RMSD = 0.23 and 0.27~\AA for $I4_1/amd$ and \textit{Cmcm}, respectively) (Fig.~\ref{FIG-1} b, d). Moreover, the fundamental structural motifs remain completely intact without any signs of decomposition, as clearly visible in the structural snapshots (see inset). These comprehensive simulations provide strong evidence for the stability of both phases under a wide range of conditions.
	
	\subsection{Magnetic ordering variations and charge compensation}
	
	\begin{figure}[hbtp]
	\centering
	\includegraphics[height=10cm]{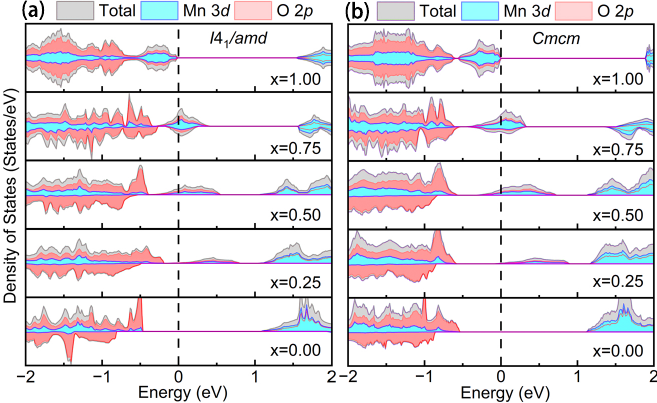}
	\caption{Total and projected DOS for (a) $I4_1/amd$ and (b) \textit{Cmcm} phases of Na\textsubscript{x}MnO\textsubscript{2} at various deintercalation levels ($x$ = 1.00, 0.75, 0.50, 0.25, 0.00). The gray shade represents the total DOS, while the blue and red shades correspond to the projected DOS of Mn 3$d$ and O 2$p$ orbitals, respectively. The Fermi level is set to zero energy and indicated by the vertical dashed line. \label{FIG-3}}
	\end{figure}
	
	The evolution of magnetic ordering and charge compensation mechanisms during sodium deintercalation was systematically investigated through density of states (DOS) calculations and Bader charge analysis at varying Na extraction levels (Figure~\ref{FIG-3} and Table~S3). The charge values presented in Table~S3 were determined by subtracting the calculated Bader charges from the nominal valence electron counts of each atomic species.
	
	Previous studies on the magnetic ordering transformation in $\beta$-NaMnO$_{2}$ and $\alpha$-NaMnO$_{2}$ during Na deintercalation have established that AFM ordering becomes energetically favorable at high Na concentrations ($>62.5\%$), while FM ordering predominates at lower Na concentrations ($<62.5\%$)\cite{AFMcankao2}. Our application of this analytical approach to the $I4_1/amd$ and \textit{Cmcm} phases yielded consistent results (Fig.S4 in Supplemental Material). Specifically, for Na$_x$MnO$_2$, the AFM configuration exhibited lower energy states at $x = 0.75$ and $1.00$, whereas FM ordering became energetically favorable at $x = 0.00$, $0.25$, and $0.50$. We attribute this magnetic behavior to the valence state transition of Mn ions: Mn$^{3+}$ (3$d^{4}$) in \ce{NaMnO2} versus Mn$^{4+}$ ($3d^3$) in \ce{Na0MnO2}. The Mn$^{3+}$ ions undergo JT distortion, while Mn$^{4+}$ remains JT-inactive. These distinct electronic configurations lead to modifications in Mn--O octahedral bond lengths, which ultimately govern the preferred magnetic ordering in the system.
	
	To verify this hypothesis, we systematically computed the average absolute magnetic moments of Mn ions across different Na concentrations. The obtained values of $3.80~\mu_{\text{B}}$, $3.60~\mu_{\text{B}}$, $3.50~\mu_{\text{B}}$, $3.35~\mu_{\text{B}}$, and $3.28~\mu_{\text{B}}$ correspond to Na concentrations of $x = 0.00$, $0.25$, $0.50$, $0.75$, and $1.00$ in Na$_x$MnO$_2$, respectively. This progressive reduction in magnetic moment with increasing Na content clearly demonstrates a valence transition from purely Mn$^{4+}$ ($3d^3$) at $x = 0.00$ to exclusively Mn$^{3+}$ (3$d^{4}$) at $x = 1.00$, with an intermediate mixed-valence state (Mn$^{3+}$/Mn$^{4+}$) at partial Na concentrations.
	
	The DOS analysis for both phases (Figure~\ref{FIG-3}) reveals complete spin degeneracy near the Fermi level at $x$ = 1.00, resulting in zero net magnetic moment. Both the $I4_1/amd$ and \textit{Cmcm} phases exhibit semiconducting behavior. Compared to the band gaps of 1.91~eV and 2.24~eV in conventional $C2/m$ and \textit{Pmmn} phases\cite{bandgap-JMCC-2025}, \textit{Cmcm} phase possesses a band gap close to that of $C2/m$, while the $I4_1/amd$ phase shows a lower band gap of $1.62$~eV, indicating enhanced electronic conductivity for battery applications. At $x$ = 0.75, Na extraction-induced structural distortions break this symmetry, creating spin-polarized states that generate a finite magnetic moment indicative of FM tendencies. Notably, the magnetic behavior at $x$ = 0.75 exhibits characteristics that deviate from conventional FM/AFM classifications. The heterogeneous distribution of Mn valence states and corresponding magnetic moment variations suggest the emergence of ferrimagnetic ordering \cite{prb1997ferrimagnetism,nature-materials-Ferrimagnetic-spintronics-2021,U3}, where localized spin alignments cannot be adequately described by uniform FM or AFM models.
	
	Bader charge analysis (Table S3 in Supplemental Material) shows the Na charge remains nearly constant at $+0.80\ e$ throughout deintercalation, confirming its ionic character. The DOS (Figure~\ref{FIG-3}) demonstrate significant Mn $3d$-O $2p$ orbital hybridization near the Fermi level, suggesting competitive redox activity between Mn and O during Na extraction. In the majority spin channel, the occupied Mn $3d$ states lie predominantly below the O $2p$ states, indicating stronger O $2p$ orbital contributions. The initial Bader charges for Mn and O are $\sim$$+1.60\ e$ and $-1.20\ e$, respectively. During the first deintercalation stage (\ce{NaMnO2} $\to$ \ce{Na_{0.5}MnO2}), both Mn ($\Delta q \approx +0.09\ e$) and O ($\Delta q \approx +0.16\ e$) undergo oxidation. The subsequent deintercalation (\ce{Na_{0.5}MnO2} $\to$ \ce{Na0MnO2}) shows reduced Mn oxidation ($\Delta q \approx +0.06\ e$) but enhanced O participation ($\Delta q \approx +0.17\ e$). These observations, combined with DOS analysis, confirm cooperative Mn-O redox behavior contributing to charge capacity. The interplay between Fermi level shifts from Na deintercalation and conduction band reduction due to magnetic ordering transformations leads to significant bandgap narrowing during desodiation. Specifically, we observe $\Delta E_g$ reductions of $1.32$~eV ($I4_1/amd$) and $1.64$~eV (\textit{Cmcm}) during the structural evolution from \ce{NaMnO2} to \ce{Na_{0.25}MnO2}, indicating improved electronic conductivity. This behavior parallels observations in \ce{Li2MnO3}\cite{Li2MnO3-2012} and \ce{Li2MnSiO4}\cite{wangshuo-Li2MnSiO4} systems. Besides, in completely de-sodiated Na$_0$MnO$_2$, the band gaps are 1.61 eV ($I4_1/amd$) and 1.56 eV (\textit{Cmcm}), respectively.
	
	\subsection{Na ion diffusion behavior and open-circuit voltage}
	
\begin{figure}[hbtp]
		\centering
		\rotatebox{0}{\includegraphics[width=1.0\textwidth]{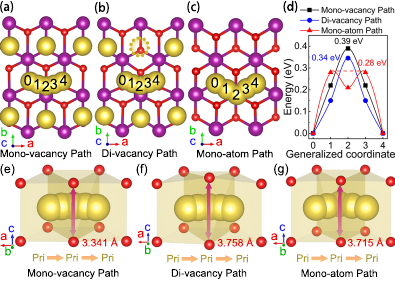}}
		\caption{(a-c) Na\textsuperscript{+} migration pathways in the \textit{Cmcm} phase: (a) mono-vacancy path, (b) di-vacancy path, and (c) mono-atom path, with (d) corresponding migration energy barriers. The black, blue, and red curves in (d) represent the energy barriers for mono-vacancy, di-vacancy, and mono-atom pathways, respectively. (e-g) Schematic illustrations of the diffusion mechanisms showing the evolution of Na\textsuperscript{+} coordination environments along each pathway. Critical O--O distances (marked by red arrows) determine the channel dimensions, with prismatic (Pri) sites explicitly labeled.}\label{FIG-4}  
	\end{figure}

\begin{figure}[htbp]
	\centering
	\rotatebox{0}{\includegraphics[width=0.95\textwidth]{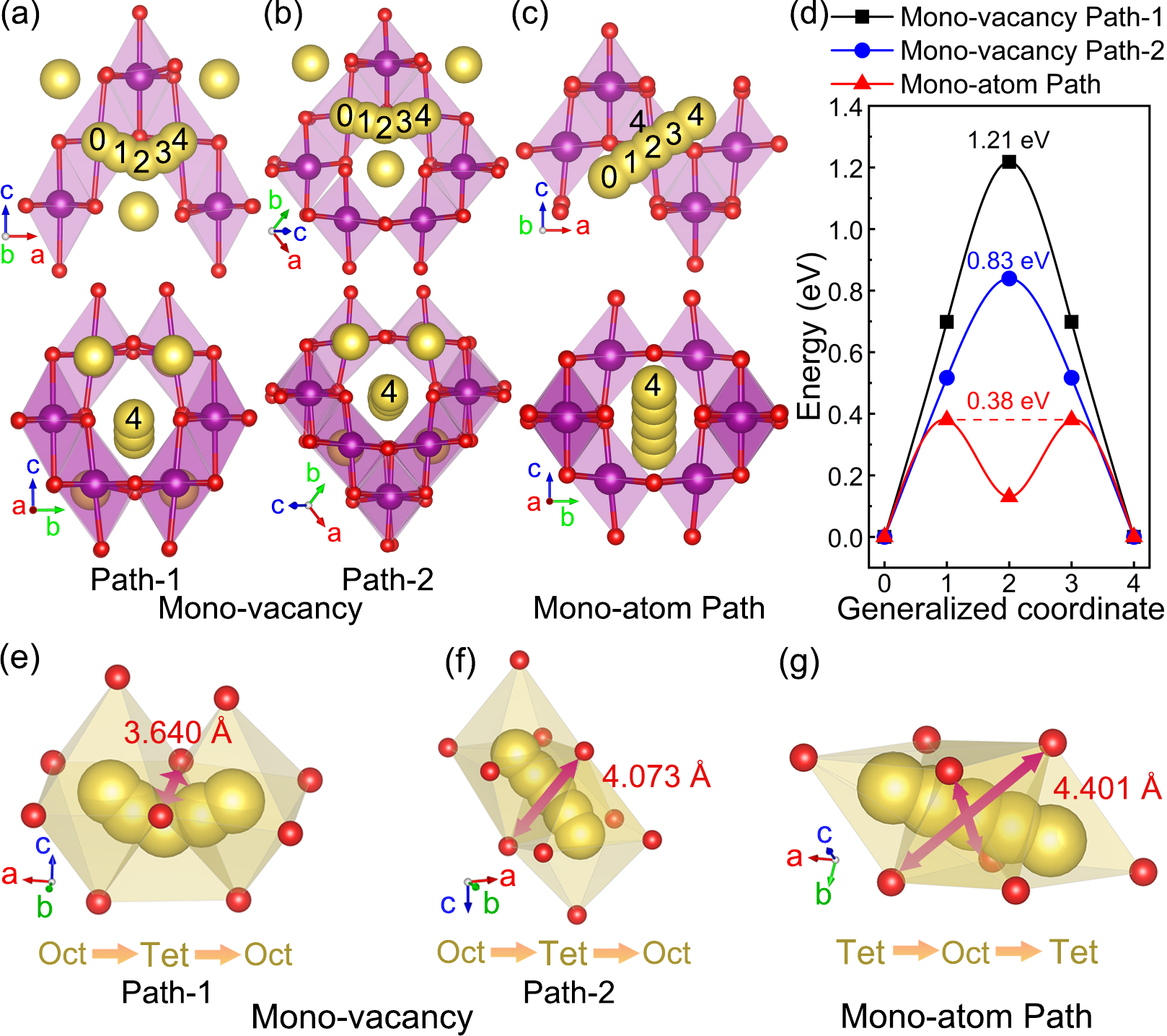}}
	\caption{(a-c) Na\textsuperscript{+} migration pathways in the $I4_1/amd$ phase: (a, b) two crystallographically distinct mono-vacancy paths (Path-1 and Path-2) and (c) mono-atom interstitial path. (d) Corresponding migration energy barriers, where black, blue, and red curves represent Path-1, Path-2, and mono-atom pathways, respectively. (e-g) Schematic representations showing the evolution of Na\textsuperscript{+} coordination environments during diffusion, with critical O--O distances (red arrows) defining the channel geometry. Key crystallographic sites are labeled: octahedral (Oct) and tetrahedral (Tet).}
		\label{FIG-5}
	\end{figure}

	To investigate the ionic migration characteristics of the \textit{Cmcm} and $I4_1/amd$ phases as cathode materials, we constructed a three-dimensional ion migration network using the vacancy diffusion method and calculated the corresponding migration energy barriers using the CI-NEB method, as shown in Fig.~\ref{FIG-4} and \ref{FIG-5}. Given that the pure zigzag-type $\beta$-NaMnO$_2$ phase has not been experimentally isolated due to its inevitable intergrowth with the thermodynamically stable $\alpha$-NaMnO$_2$ ($C2/m$) phase under ambient conditions \cite{AFMcankao3-compare-daixicankao,U2-compare,parant1971quelques,beita}, our comparative analysis focuses exclusively on the $C2/m$ polymorph. The calculated properties of this reference phase show excellent agreement with prior theoretical results reported by Zhu \textit{et al}.\cite{U1-cengzhuangnengleicankao}, thereby validating our computational methodology. Our mono-vacancy model for the $C2/m$ structure at high Na concentration yields a migration barrier of 0.70 eV, in excellent agreement with their reported value of 0.74 eV, thereby validating our methodology. For low Na concentrations, the mono-atom approach predicts a significantly reduced barrier of 0.42 eV, as listed in Table~\ref{table1}.
	
	\begin{table*}[hbtp]
		\centering
		\caption{Migration energy barriers and electrochemical properties of NaMnO\textsubscript{2} polymorphs}
		\label{table1}
		\begin{ruledtabular}
			\begin{tabular}{lccc}
				Phases & \multicolumn{2}{c}{Migration barrier (eV)} & Open-circuit voltage (V) \\
				\cmidrule(lr){2-3} \cmidrule(lr){4-4}
				& \begin{tabular}{@{}c@{}}High Na concentration \\ (mono-vacancy)\end{tabular} & \begin{tabular}{@{}c@{}}Low Na concentration \\ (mono-atom)\end{tabular} & \\
				\midrule
				$C2/m$ & 0.74 \cite{U1-cengzhuangnengleicankao} & 0.42 & 2.88 (2.45--3.50) \cite{toumar2015vacancy} \\
				\textit{Cmcm} (this work) & 0.39 & 0.28 & 3.19 (2.84--3.56) \\
				$I4_1/amd$ (this work) & 0.83 (Path-2) & 0.38 & 2.94 (2.54--3.38) \\
				& 1.21 (Path-1) & & \\
			\end{tabular}
		\end{ruledtabular}
	\end{table*}
	
	Our DFT calculations reveal that the \textit{Cmcm} phase exhibits remarkably lower Na\textsuperscript{+} migration barriers, as shown in Fig.~\ref{FIG-4}. At high Na concentration, the mono-vacancy configuration yields a migration barrier of 0.39~eV. Recognizing the critical role of vacancy concentration in ionic transport \cite{vacancy-effect}, we systematically investigated both di-vacancy and low-concentration regimes. For the di-vacancy configuration, illustrated in Fig.~\ref{FIG-4}(b), where an additional vacancy is introduced above the primary $b$-axis migration channel, the activation energy reduces to 0.34~eV---a 0.05~eV decrease attributed to channel widening, as evidenced by the increased O--O separation from 3.341~\AA\ to 3.758~\AA\, as presented in Fig.~\ref{FIG-4}(e-f).
	
	At low Na concentration, the mono-atom model reveals further kinetic enhancement: The absence of interlayer Na\textsuperscript{+} repulsion stabilizes an intermediate state at the Na2 site (0.21~eV above the initial state), while the O--O distance expands to 3.715~\AA, as shown in Fig.~\ref{FIG-4}(g). This geometric adjustment reduces the overall barrier to 0.28~eV, suggesting superior ionic conductivity in the \textit{Cmcm} phase compared to conventional $C2/m$ across all Na concentrations, see Table~\ref{table1}. Notably, all three pathways maintain a consistent prismatic-prismatic-prismatic coordination over 6.5~\AA, despite their differing vacancy contexts. This robust structural motif, combined with the observed strong correlation between local channel geometry and migration barriers, highlights the \textit{Cmcm} phase's unique ability to adapt its diffusion kinetics to vacancy concentration---a crucial design principle for optimized Na-ion conductors.
	
	\begin{figure}[hbtp]
	\centering
	\rotatebox{0}{\includegraphics[width=1.0\textwidth]{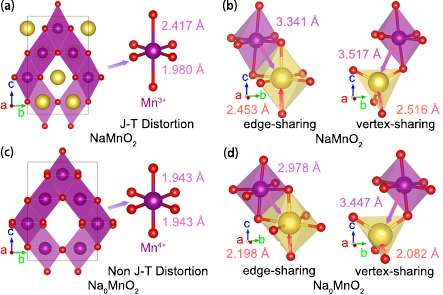}}
	\caption{Crystal structures of (a) fully sodiated NaMnO\textsubscript{2} and (c) fully desodiated Na\textsubscript{0}MnO\textsubscript{2} in the $I4_1/amd$ phase. MnO\textsubscript{6} octahedra with different valence states and their Mn--O bond lengths are labeled with purple arrows. (b, d) Coordination environments of octahedral/tetrahedral NaO\textsubscript{6}/NaO\textsubscript{4} with nearest MnO\textsubscript{6} octahedra and Na--Mn bond lengths. \label{FIG-6}}
	\end{figure}
	
	In the $I4_1/amd$ phase, our analysis reveals two crystallographically distinct mono-vacancy migration pathways, as illustrated in Fig.~\ref{FIG-5}(a,b). Both Path-1 and Path-2 exhibit an octahedral-tetrahedral-octahedral hopping sequence, but with significantly different activation energies of 1.21 eV and 0.83 eV, as shown in Fig.~\ref{FIG-5}(d). This 0.38 eV difference primarily originates from the expanded diffusion channel geometry in Path-2, where the critical O--O separation measures 4.073~\AA\ compared to 3.640~\AA\ in Path-1, as clearly shown in Fig.~\ref{FIG-5}(e,f). To elucidate the low-concentration migration behavior, we performed additional calculations using a mono-atom approach, as illustrated in Fig.~\ref{FIG-5}(c). In this configuration, isolated Na\textsuperscript{+} ions preferentially occupy tetrahedral sites that vertex-share with MnO\textsubscript{6} octahedra. The migration pathway proceeds through an octahedral intermediate state with a modest energy increase of 0.13~eV relative to the initial position, yielding an exceptionally low activation barrier of 0.38~eV, which demonstrates that at low Na concentration, the $I4_1/amd$ phase exhibits superior ionic conductivity compared to the conventional layered $C2/m$ phase, as shown in Table \ref{table1}. This substantial reduction in migration energy correlates with the expanded geometry of the diffusion channel, where the critical O--O separation measures 4.401~\AA---significantly wider than the corresponding distances of 3.640~\AA (Path-1) and 4.073~\AA (Path-2) observed in the mono-vacancy cases (Fig.~\ref{FIG-5}(g)). The direct comparison between these configurations provides compelling evidence for the structure-property relationship governing Na\textsuperscript{+} mobility in $I4_1/amd$ phase.
	
	The observed tetrahedral coordination of Na$^+$ in $I4_1/amd$ phase appears to contradict Pauling's rules for cation coordination in oxides \cite{Pauling-rules}, which typically favor octahedral or higher coordination environments for Na$^+$\cite{PvsO}. To resolve this apparent discrepancy, we performed detailed structural analysis of both fully sodiated NaMnO$_2$ and completely desodiated Na$_0$MnO$_2$, as shown in Fig.~\ref{FIG-6}(a, c). The desodiation process induces anisotropic lattice contraction: the calculated in-plane lattice parameters $a$ and $b$ decrease by 3.3\% (from 6.027 \AA\ to 5.827 \AA), while the out-of-plane parameter $c$ shows a substantially larger contraction of 15.3\% (from 9.702 \AA\ to 8.222 \AA). This pronounced anisotropic distortion stems from the valence transition of Mn ions from JT active Mn$^{3+}$ (3d$^4$) to non-JT active Mn$^{4+}$ (3d$^3$), as evidenced by the evolution of MnO$_6$ octahedral geometry: the axial Mn-O bonds in NaMnO$_2$ (2.417 \AA) are significantly longer than the equatorial bonds (1.980 \AA), while Na$_0$MnO$_2$ exhibits regular octahedra with an identical bond length of 1.943 \AA.
	
	Further analysis of local coordination environments shows distinct behavior for octahedral and tetrahedral sites, see Fig.~\ref{FIG-6}(b, d). In octahedral sites, the Na-Mn distance contracts by 0.363 \AA\ (from 3.341 \AA\ to 2.978 \AA) while the Na-O distance decreases by only 0.255 \AA\ (2.453 \AA\ to 2.198 \AA), leading to enhanced Na-Mn repulsion. Conversely, in tetrahedral sites, the Na-O bond shortens significantly by 0.434 \AA\ (2.516 \AA\ to 2.082 \AA) compared to a modest 0.070 \AA\ reduction in Na-Mn distance (3.517 \AA\ to 3.447 \AA), resulting in stronger Na-O attraction. This contrasting behavior creates an energetic preference for tetrahedral coordination at low Na concentrations, where the enhanced Na-O bonding compensates for the deviation from Pauling's rules. While this coordination-dependent energetics suggests Na$^+$ site preference varies with concentration, the stability of tetrahedral NaO$_4$ coordination under operational conditions requires further investigation through combined theoretical and experimental studies to assess its implications for practical battery applications.
	
	\begin{figure}[htbp]
		\centering
		\rotatebox{0}{\includegraphics[width=0.85\textwidth]{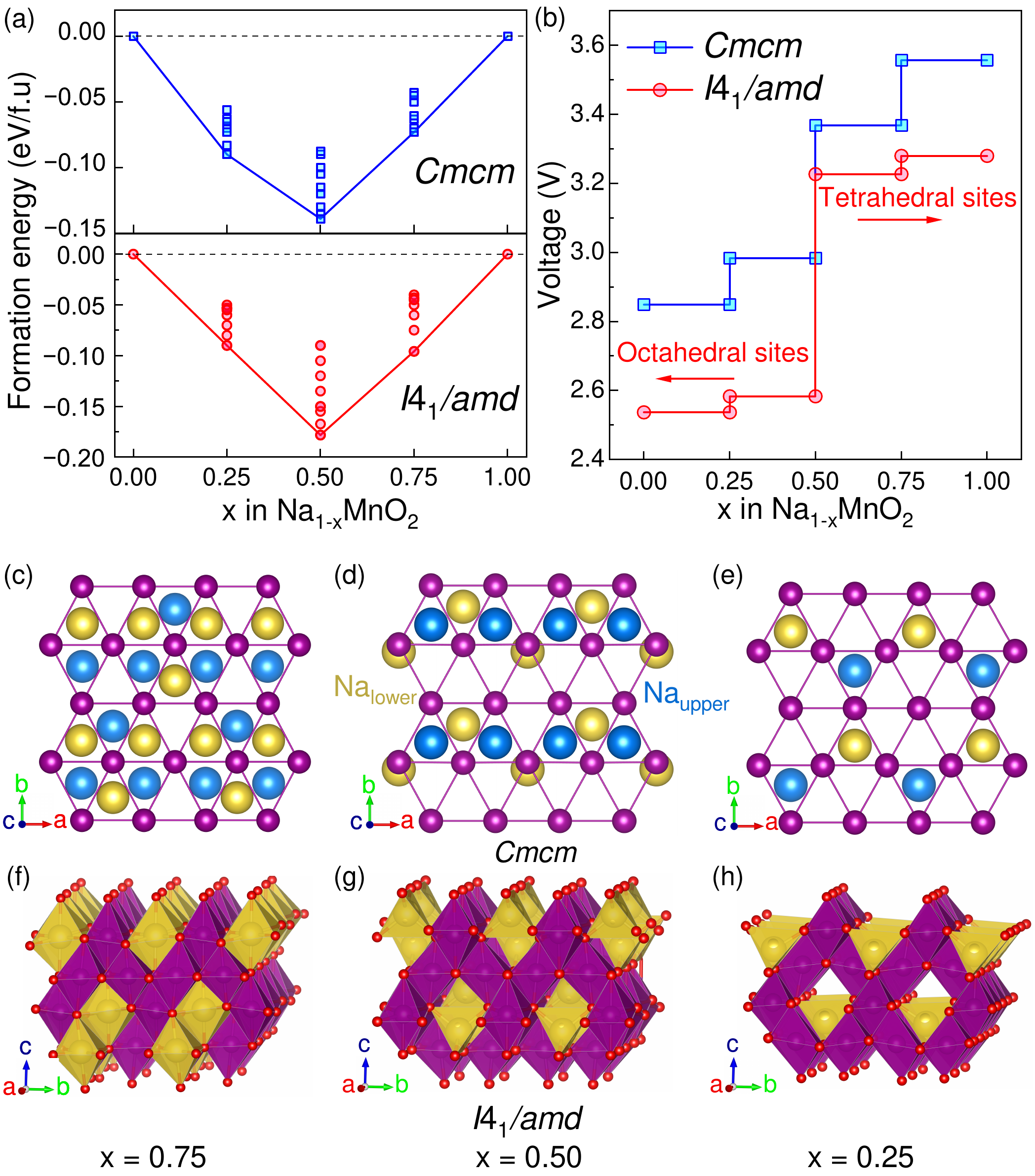}}
		\caption{(a) Formation energy as a function of sodium concentration $x$ in Na$_{1-x}$MnO$_2$ for the $I4_1/amd$ (red circles) and \textit{Cmcm} (blue squares) phases, calculated with respect to the pristine compounds ($x$ = 0 and 1). (b) Corresponding equilibrium voltage profiles versus Na/Na$^+$, where error bars indicate the hysteresis between charge and discharge cycles. (c-e) Atomic structures of the lowest-energy configurations at (c) $x$ = 0.75, (d) $x$ = 0.50, and (e) $x$ = 0.25, illustrating the distinct Na$_\mathrm{upper}$ (yellow) and Na$_\mathrm{lower}$ (blue) sites in the \textit{Cmcm} phase, and (f-h) the evolution from octahedral to tetrahedral Na coordination in the $I4_1/amd$ phase with decreasing Na concentration ($x$) is depicted. A polyhedral representation is employed for the $I4_1/amd$ structures to more clearly illustrate the transition in Na coordination geometry. \label{FIG-7}}
	\end{figure}
	
	The OCV is a critical descriptor for evaluating the performance of SIB cathodes. Within the framework of density functional theory, the OCV profile can be estimated from total energy differences, neglecting entropic, volumetric, and pressure contributions. The average voltage between sodium concentrations $x_1 < x < x_2$ is given by:
	\begin{equation}
		V \approx \frac{E_{\text{Na}_{x_1} \text{MnO}_2} - E_{\text{Na}_{x_2} \text{MnO}_2} - (x_1 - x_2) E_{\text{Na}}}{x_1 - x_2}
	\end{equation}
	where \( E_{\text{Na}_{x_1} \text{MnO}_2} \) and \( E_{\text{Na}_{x_2} \text{MnO}_2} \) are the total energies per formula unit (f.u.) of Na$_{x}$MnO$_2$ at concentrations $x_1$ and $x_2$, respectively, and \( E_{\text{Na}} \) is the energy per atom of body-centered cubic (bcc) sodium\cite{bcc-Na-Nature-2007,bcc-Na-prb-2020}.
	
	To account for Na/vacancy ordering, ten symmetry-inequivalent configurations were constructed for each intermediate concentration ($x$ = 0.25, 0.5, and 0.75). The configuration with the lowest electrostatic energy was selected for total energy calculations. The corresponding formation energy, computed with respect to NaMnO$_2$ and MnO$_2$ end members, is defined as:
	\begin{equation}
		E_f(x) = E_{\text{Na}_x\text{MnO}_2} - (1 - x)E_{\text{MnO}_2} - xE_{\text{NaMnO}_2}
	\end{equation}
	
	The calculated formation energies are presented in Fig.~\ref{FIG-7}(a). In both the \textit{Cmcm} and $I4_1/amd$ phases, three intermediate compositions ($x = 0.25$, $0.5$, and $0.75$) are thermodynamically stable, exhibiting negative formation energies with respect to the reference states. The corresponding equilibrium voltage profiles versus Na/Na$^+$ are shown in Fig.~\ref{FIG-7}(b). For the layered \textit{Cmcm} phase, four voltage plateaus are observed at 2.84, 2.98, 3.37, and 3.56~V, with an average voltage of 3.19~V, demonstrating a higher average voltage plateau than the conventional $C2/m$ phase (2.88 eV) calculated by Toumar \textit{et al.}\cite{toumar2015vacancy}, as listed in Table \ref{table1}. The voltage steps originate from the distinct Na ordering patterns and resulting structural stabilization at specific Na contents.
	
	The atomic configurations of the lowest-energy structures at $x$ = 0.75, 0.50, and 0.25 are illustrated in Fig.~\ref{FIG-7}(c-e). At $x$ = 0.75 and  0.25, Na$^+$ ions adopt triangular arrangements within the sodium layers and form a rhombic motif between adjacent layers (viewed along the $c$-axis), with Na$^+$ ions centered within Mn-constructed triangles. In these configurations, NaO$_6$ and MnO$_6$ octahedra are edge-sharing, consistent with the findings reported by Luong \textit{et al.}~\cite{Cmcm-DFT-pccp-2020}. Notably, our calculations yield lower total energies for these ordered structures compared to their reported configurations. At $x$ = 0.5, Na$^+$ ions exhibit a mixed coordination environment, interacting with MnO$_6$ octahedra via edge-sharing and face-sharing at a 2:1 ratio, as shown in Fig.~\ref{FIG-7}(d). The ions are arranged in a zigzag pattern along the $a$-axis, occupying both the centers of Mn triangles and positions directly below Mn ions. The emergence of face-sharing configurations can be attributed to competing intralayer Na$^+$--Na$^+$ repulsion and interlayer Na$^+$--Mn$^{3+}$ repulsion, similar to the behavior observed in doped systems studied by Daubner \textit{et al.}~\cite{npj-2024}.
	
	In the $I4_1/amd$ phase, all Na ions preferentially occupy tetrahedral sites at $x = 0.5$ and $x = 0.75$, as shown in Fig.~\ref{FIG-7}(f-h), analogous to the behavior in spinel \ce{LiMn2O4} where Li ions migrate from tetrahedral to octahedral sites upon delithiation, resulting in a steep voltage increase. A similar trend is observed in our system: upon Na extraction from octahedral sites, a sharp voltage rise occurs, as shown in Fig.~\ref{FIG-7}(b). The voltage profile reveals four distinct plateaus at 2.54, 2.59, 3.24, and 3.38~V, with an average value of 2.94~V, also demonstrating a higher average voltage plateau than the conventional $C2/m$ phase (2.88 eV), as presented in Table \ref{table1}. However, unlike \ce{LiMn2O4}, the final de-sodiation step does not reach voltages above 4~V, which can be attributed to the larger ionic radius of \ce{Na^+}, leading to less favorable occupation of tetrahedral sites and lower structural stability at deep de-sodiation.
	
	For oxide-based cathode materials, higher oxidation states may lead to oxygen release, resulting in irreversible capacity decay and decreased cycling stability\cite{npj-ceder-2016,wangshuo-LiCoO2,oxygen-stability-JMCA-2022}. To examine the oxygen stability in both \textit{Cmcm} and $I4_1/amd$ phases at the high-voltage region, we evaluated the change of Gibbs free energy ($\Delta G$) for the formation of oxygen vacancies in Na$_{1-x}$MnO$_2$ at $x$ = 1.0. $\Delta G$ values were obtained based on the following equation:
	\begin{equation}
		\Delta G = \Delta H -T\Delta S 
	\end{equation}
	where $\Delta H$ is the enthalpy of oxygen vacancy formation at $x$ = 1.0 and -T$\Delta S$ is the entropy of gas phase O$_{2}$ under standard conditions at 298.15 K. The $\Delta H$ was obtained of the following reaction:
	\begin{equation}
		\text{Na}_0\text{MnO}_2(\text{s}) \rightarrow \text{Na}_0\text{MnO}_{2-y}(\text{s}) + \frac{y}{2} \text{O}_2(\text{g})
	\end{equation}
	The change of enthalpy is calculated as:
	\begin{equation}
		\Delta H = E_{\text{Na}_0\text{MnO}_{2-y}} + \frac{y}{2} E_{\text{O}_2} - E_{\text{Na}_0\text{MnO}_2}
	\end{equation}
	where $E$ denotes the total energy per f.u. To construct the oxygen-deficient model, the oxygen atom with the smallest Bader atomic charge was removed from the cell, since an O$^{2-}$ ion that has lost more electrons exhibits a stronger tendency to be oxidized to O$_{2}$ gas.
	Accurate calculation of the \textit{ab initio} energy of O$_2$ is challenging because the DFT-GGA method introduces significant errors for the O$_2$ molecule, leading to an overestimation of its binding energy in DFT calculations\cite{DFT-bad-for-O2-1,DFT-bad-for-O2-2}. Here, the more accurate \textit{ab initio} energies of H$_2$O and H$_2$, together with the experimental formation enthalpy of water\cite{experimentally-water}, are adopted to obtain the energy of the O$_2$ molecule as follows\cite{Equation-JCP-2010}:
	\begin{equation}
		E(\text{O}_2) = 2E_{\text{DFT}}(\text{H}_2\text{O}) - 2E_{\text{DFT}}(\text{H}_2) - 2\Delta E_{\text{exp}}(\text{H}_2\text{O})
	\end{equation}
	
	The calculated $\Delta G$ values for oxygen vacancy formation are +1.98~eV and +1.89~eV for the \textit{Cmcm} and $I4_1/amd$ phases, respectively. These positive values indicate that the formation of an oxygen vacancy is thermodynamically unfavorable in both phases, even under fully desodiation conditions.
	
	\section{CONCLUSION}
	Using \textit{ab initio} EA and first-principles calculations, we identified two metastable NaMnO\textsubscript{2} polymorphs, \textit{Cmcm} and $I4_1/amd$, as promising cathode candidates for SIBs. Both phases exhibit excellent thermodynamic stability, lying within 35~meV/atom of the ground-state \textit{Pmmn} phase over 0--50~GPa, indicating experimental accessibility via high-pressure synthesis and quenching. Phonon dispersion and \textit{ab initio} molecular dynamics simulations further confirm their dynamic and thermal stability under operating conditions. During desodiation, a Jahn--Teller-driven transition from antiferromagnetic to ferromagnetic ordering enhances Mn--O hybridization, reduces electronic bandgaps, and improves electronic conductivity.
	
	Remarkably, the newly identified \textit{Cmcm} phase demonstrates superior Na$^+$ diffusion kinetics compared to the conventional $C2/m$ phase, exhibiting significantly lower migration barriers (0.39~eV \textit{versus} 0.74~eV at high Na concentration; 0.28~eV \textit{versus} 0.42~eV at low Na concentration) while delivering a higher average voltage (3.19~V \textit{versus} 2.88~V). The $I4_1/amd$ phase shows concentration-dependent migration behavior with competitive barriers (0.38~eV compared to $C2/m$'s 0.42~eV at low concentration) and maintains comparable voltage characteristics (2.94~V average). Both structures maintain positive formation energies for oxygen vacancies even under full desodiation, indicating good oxygen retention and cycling stability. These favorable characteristics---fast ion transport, stable voltage profiles, efficient charge compensation, and robust oxygen stability---highlight the potential of \textit{Cmcm} and $I4_1/amd$ as promising candidates for cathode materials beyond conventional NaMnO\textsubscript{2}.
	
	\begin{acknowledgments}
		This work was supported by the Innovation Capability Improvement Project of Hebei province (Grant No. 22567605H). The numerical calculations in this paper have been done on the supercomputing system in the High Performance Computing Center of Yanshan University.
	\end{acknowledgments}
	

\begin{thebibliography}{71}%
		\makeatletter
		\providecommand \@ifxundefined [1]{%
			\@ifx{#1\undefined}
		}%
		\providecommand \@ifnum [1]{%
			\ifnum #1\expandafter \@firstoftwo
			\else \expandafter \@secondoftwo
			\fi
		}%
		\providecommand \@ifx [1]{%
			\ifx #1\expandafter \@firstoftwo
			\else \expandafter \@secondoftwo
			\fi
		}%
		\providecommand \natexlab [1]{#1}%
		\providecommand \enquote  [1]{``#1''}%
		\providecommand \bibnamefont  [1]{#1}%
		\providecommand \bibfnamefont [1]{#1}%
		\providecommand \citenamefont [1]{#1}%
		\providecommand \href@noop [0]{\@secondoftwo}%
		\providecommand \href [0]{\begingroup \@sanitize@url \@href}%
		\providecommand \@href[1]{\@@startlink{#1}\@@href}%
		\providecommand \@@href[1]{\endgroup#1\@@endlink}%
		\providecommand \@sanitize@url [0]{\catcode `\\12\catcode `\$12\catcode
			`\&12\catcode `\#12\catcode `\^12\catcode `\_12\catcode `\%12\relax}%
		\providecommand \@@startlink[1]{}%
		\providecommand \@@endlink[0]{}%
		\providecommand \url  [0]{\begingroup\@sanitize@url \@url }%
		\providecommand \@url [1]{\endgroup\@href {#1}{\urlprefix }}%
		\providecommand \urlprefix  [0]{URL }%
		\providecommand \Eprint [0]{\href }%
		\providecommand \doibase [0]{http://dx.doi.org/}%
		\providecommand \selectlanguage [0]{\@gobble}%
		\providecommand \bibinfo  [0]{\@secondoftwo}%
		\providecommand \bibfield  [0]{\@secondoftwo}%
		\providecommand \translation [1]{[#1]}%
		\providecommand \BibitemOpen [0]{}%
		\providecommand \bibitemStop [0]{}%
		\providecommand \bibitemNoStop [0]{.\EOS\space}%
		\providecommand \EOS [0]{\spacefactor3000\relax}%
		\providecommand \BibitemShut  [1]{\csname bibitem#1\endcsname}%
		\let\auto@bib@innerbib\@empty
		\bibitem [{\citenamefont {Whittingham}(2014)}]{Li-limited}%
		\BibitemOpen
		\bibfield  {author} {\bibinfo {author} {\bibfnamefont {M.~S.}\ \bibnamefont
				{Whittingham}},\ }\href@noop {} {\bibfield  {journal} {\bibinfo  {journal}
				{Chem. Rev.}\ }\textbf {\bibinfo {volume} {114}},\ \bibinfo {pages} {11414}
			(\bibinfo {year} {2014})}\BibitemShut {NoStop}%
		\bibitem [{\citenamefont {Fan}\ \emph {et~al.}(2020)\citenamefont {Fan},
			\citenamefont {Li}, \citenamefont {Wang}, \citenamefont {Lin}, \citenamefont
			{Huang}, \citenamefont {Yao}, \citenamefont {Chen},\ and\ \citenamefont
			{Wu}}]{Li-limited-2020}%
		\BibitemOpen
		\bibfield  {author} {\bibinfo {author} {\bibfnamefont {E.}~\bibnamefont
				{Fan}}, \bibinfo {author} {\bibfnamefont {L.}~\bibnamefont {Li}}, \bibinfo
			{author} {\bibfnamefont {Z.}~\bibnamefont {Wang}}, \bibinfo {author}
			{\bibfnamefont {J.}~\bibnamefont {Lin}}, \bibinfo {author} {\bibfnamefont
				{Y.}~\bibnamefont {Huang}}, \bibinfo {author} {\bibfnamefont
				{Y.}~\bibnamefont {Yao}}, \bibinfo {author} {\bibfnamefont {R.}~\bibnamefont
				{Chen}}, \ and\ \bibinfo {author} {\bibfnamefont {F.}~\bibnamefont {Wu}},\
		}\href@noop {} {\bibfield  {journal} {\bibinfo  {journal} {Chem. Rev.}\
			}\textbf {\bibinfo {volume} {120}},\ \bibinfo {pages} {7020} (\bibinfo {year}
			{2020})}\BibitemShut {NoStop}%
		\bibitem [{\citenamefont {Huang}\ \emph {et~al.}(2022)\citenamefont {Huang},
			\citenamefont {Boles},\ and\ \citenamefont {Tarascon}}]{Li-limited-2022}%
		\BibitemOpen
		\bibfield  {author} {\bibinfo {author} {\bibfnamefont {J.}~\bibnamefont
				{Huang}}, \bibinfo {author} {\bibfnamefont {S.~T.}\ \bibnamefont {Boles}}, \
			and\ \bibinfo {author} {\bibfnamefont {J.-M.}\ \bibnamefont {Tarascon}},\
		}\href@noop {} {\bibfield  {journal} {\bibinfo  {journal} {Nat. Sustain.}\
			}\textbf {\bibinfo {volume} {5}},\ \bibinfo {pages} {194} (\bibinfo {year}
			{2022})}\BibitemShut {NoStop}%
		\bibitem [{\citenamefont {Ramasubramanian}\ \emph {et~al.}(2024)\citenamefont
			{Ramasubramanian}, \citenamefont {Ling}, \citenamefont {Jose},\ and\
			\citenamefont {Ramakrishna}}]{Li-limited-2024}%
		\BibitemOpen
		\bibfield  {author} {\bibinfo {author} {\bibfnamefont {B.}~\bibnamefont
				{Ramasubramanian}}, \bibinfo {author} {\bibfnamefont {J.}~\bibnamefont
				{Ling}}, \bibinfo {author} {\bibfnamefont {R.}~\bibnamefont {Jose}}, \ and\
			\bibinfo {author} {\bibfnamefont {S.}~\bibnamefont {Ramakrishna}},\
		}\href@noop {} {\bibfield  {journal} {\bibinfo  {journal} {Cell Rep. Phys.
					Sci.}\ }\textbf {\bibinfo {volume} {5}},\ \bibinfo {pages} {102032} (\bibinfo
			{year} {2024})}\BibitemShut {NoStop}%
		\bibitem [{\citenamefont {Slater}\ \emph {et~al.}(2013)\citenamefont {Slater},
			\citenamefont {Kim}, \citenamefont {Lee},\ and\ \citenamefont
			{Johnson}}]{slater2013sodium}%
		\BibitemOpen
		\bibfield  {author} {\bibinfo {author} {\bibfnamefont {M.~D.}\ \bibnamefont
				{Slater}}, \bibinfo {author} {\bibfnamefont {D.}~\bibnamefont {Kim}},
			\bibinfo {author} {\bibfnamefont {E.}~\bibnamefont {Lee}}, \ and\ \bibinfo
			{author} {\bibfnamefont {C.~S.}\ \bibnamefont {Johnson}},\ }\href@noop {}
		{\bibfield  {journal} {\bibinfo  {journal} {Adv. Funct. Mater.}\ }\textbf
			{\bibinfo {volume} {23}},\ \bibinfo {pages} {947} (\bibinfo {year}
			{2013})}\BibitemShut {NoStop}%
		\bibitem [{\citenamefont {Yabuuchi}\ \emph {et~al.}(2014)\citenamefont
			{Yabuuchi}, \citenamefont {Kubota}, \citenamefont {Dahbi},\ and\
			\citenamefont {Komaba}}]{yabuuchi2014research}%
		\BibitemOpen
		\bibfield  {author} {\bibinfo {author} {\bibfnamefont {N.}~\bibnamefont
				{Yabuuchi}}, \bibinfo {author} {\bibfnamefont {K.}~\bibnamefont {Kubota}},
			\bibinfo {author} {\bibfnamefont {M.}~\bibnamefont {Dahbi}}, \ and\ \bibinfo
			{author} {\bibfnamefont {S.}~\bibnamefont {Komaba}},\ }\href@noop {}
		{\bibfield  {journal} {\bibinfo  {journal} {Chem. Rev.}\ }\textbf {\bibinfo
				{volume} {114}},\ \bibinfo {pages} {11636} (\bibinfo {year}
			{2014})}\BibitemShut {NoStop}%
		\bibitem [{\citenamefont {Hwang}\ \emph {et~al.}(2017)\citenamefont {Hwang},
			\citenamefont {Myung},\ and\ \citenamefont {Sun}}]{hwang2017sodium}%
		\BibitemOpen
		\bibfield  {author} {\bibinfo {author} {\bibfnamefont {J.-Y.}\ \bibnamefont
				{Hwang}}, \bibinfo {author} {\bibfnamefont {S.-T.}\ \bibnamefont {Myung}}, \
			and\ \bibinfo {author} {\bibfnamefont {Y.-K.}\ \bibnamefont {Sun}},\
		}\href@noop {} {\bibfield  {journal} {\bibinfo  {journal} {Chem. Soc. Rev.}\
			}\textbf {\bibinfo {volume} {46}},\ \bibinfo {pages} {3529} (\bibinfo {year}
			{2017})}\BibitemShut {NoStop}%
		\bibitem [{\citenamefont {Nayak}\ \emph {et~al.}(2018)\citenamefont {Nayak},
			\citenamefont {Yang}, \citenamefont {Brehm},\ and\ \citenamefont
			{Adelhelm}}]{nayak2018lithium}%
		\BibitemOpen
		\bibfield  {author} {\bibinfo {author} {\bibfnamefont {P.~K.}\ \bibnamefont
				{Nayak}}, \bibinfo {author} {\bibfnamefont {L.}~\bibnamefont {Yang}},
			\bibinfo {author} {\bibfnamefont {W.}~\bibnamefont {Brehm}}, \ and\ \bibinfo
			{author} {\bibfnamefont {P.}~\bibnamefont {Adelhelm}},\ }\href@noop {}
		{\bibfield  {journal} {\bibinfo  {journal} {Angew. Chem. Int. Ed.}\ }\textbf
			{\bibinfo {volume} {57}},\ \bibinfo {pages} {102} (\bibinfo {year}
			{2018})}\BibitemShut {NoStop}%
		\bibitem [{\citenamefont {Vaalma}\ \emph {et~al.}(2018)\citenamefont {Vaalma},
			\citenamefont {Buchholz}, \citenamefont {Weil},\ and\ \citenamefont
			{Passerini}}]{vaalma2018cost}%
		\BibitemOpen
		\bibfield  {author} {\bibinfo {author} {\bibfnamefont {C.}~\bibnamefont
				{Vaalma}}, \bibinfo {author} {\bibfnamefont {D.}~\bibnamefont {Buchholz}},
			\bibinfo {author} {\bibfnamefont {M.}~\bibnamefont {Weil}}, \ and\ \bibinfo
			{author} {\bibfnamefont {S.}~\bibnamefont {Passerini}},\ }\href@noop {}
		{\bibfield  {journal} {\bibinfo  {journal} {Nat. Rev. Mater.}\ }\textbf
			{\bibinfo {volume} {3}},\ \bibinfo {pages} {1} (\bibinfo {year}
			{2018})}\BibitemShut {NoStop}%
		\bibitem [{\citenamefont {Zhao}\ \emph {et~al.}(2023)\citenamefont {Zhao},
			\citenamefont {Kang}, \citenamefont {Wozny}, \citenamefont {Lu},
			\citenamefont {Du}, \citenamefont {Li}, \citenamefont {Li}, \citenamefont
			{Kang}, \citenamefont {Tavajohi},\ and\ \citenamefont
			{Li}}]{zhao2023recycling}%
		\BibitemOpen
		\bibfield  {author} {\bibinfo {author} {\bibfnamefont {Y.}~\bibnamefont
				{Zhao}}, \bibinfo {author} {\bibfnamefont {Y.}~\bibnamefont {Kang}}, \bibinfo
			{author} {\bibfnamefont {J.}~\bibnamefont {Wozny}}, \bibinfo {author}
			{\bibfnamefont {J.}~\bibnamefont {Lu}}, \bibinfo {author} {\bibfnamefont
				{H.}~\bibnamefont {Du}}, \bibinfo {author} {\bibfnamefont {C.}~\bibnamefont
				{Li}}, \bibinfo {author} {\bibfnamefont {T.}~\bibnamefont {Li}}, \bibinfo
			{author} {\bibfnamefont {F.}~\bibnamefont {Kang}}, \bibinfo {author}
			{\bibfnamefont {N.}~\bibnamefont {Tavajohi}}, \ and\ \bibinfo {author}
			{\bibfnamefont {B.}~\bibnamefont {Li}},\ }\href@noop {} {\bibfield  {journal}
			{\bibinfo  {journal} {Nat. Rev. Mater.}\ }\textbf {\bibinfo {volume} {8}},\
			\bibinfo {pages} {623} (\bibinfo {year} {2023})}\BibitemShut {NoStop}%
		\bibitem [{\citenamefont {Han}\ \emph {et~al.}(2015)\citenamefont {Han},
			\citenamefont {Gonzalo}, \citenamefont {Singh},\ and\ \citenamefont
			{Rojo}}]{NaxTMO2}%
		\BibitemOpen
		\bibfield  {author} {\bibinfo {author} {\bibfnamefont {M.~H.}\ \bibnamefont
				{Han}}, \bibinfo {author} {\bibfnamefont {E.}~\bibnamefont {Gonzalo}},
			\bibinfo {author} {\bibfnamefont {G.}~\bibnamefont {Singh}}, \ and\ \bibinfo
			{author} {\bibfnamefont {T.}~\bibnamefont {Rojo}},\ }\href@noop {} {\bibfield
			{journal} {\bibinfo  {journal} {Energy Environ. Sci.}\ }\textbf {\bibinfo
				{volume} {8}},\ \bibinfo {pages} {81} (\bibinfo {year} {2015})}\BibitemShut
		{NoStop}%
		\bibitem [{\citenamefont {Toumar}\ \emph {et~al.}(2015)\citenamefont {Toumar},
			\citenamefont {Ong}, \citenamefont {Richards}, \citenamefont {Dacek},\ and\
			\citenamefont {Ceder}}]{toumar2015vacancy}%
		\BibitemOpen
		\bibfield  {author} {\bibinfo {author} {\bibfnamefont {A.~J.}\ \bibnamefont
				{Toumar}}, \bibinfo {author} {\bibfnamefont {S.~P.}\ \bibnamefont {Ong}},
			\bibinfo {author} {\bibfnamefont {W.~D.}\ \bibnamefont {Richards}}, \bibinfo
			{author} {\bibfnamefont {S.}~\bibnamefont {Dacek}}, \ and\ \bibinfo {author}
			{\bibfnamefont {G.}~\bibnamefont {Ceder}},\ }\href@noop {} {\bibfield
			{journal} {\bibinfo  {journal} {Phys. Rev. Appl.}\ }\textbf {\bibinfo
				{volume} {4}},\ \bibinfo {pages} {064002} (\bibinfo {year}
			{2015})}\BibitemShut {NoStop}%
		\bibitem [{\citenamefont {Wang}\ \emph
			{et~al.}(2019{\natexlab{a}})\citenamefont {Wang}, \citenamefont {Sun},
			\citenamefont {Wang},\ and\ \citenamefont {Zhang}}]{wang2019ni}%
		\BibitemOpen
		\bibfield  {author} {\bibinfo {author} {\bibfnamefont {S.}~\bibnamefont
				{Wang}}, \bibinfo {author} {\bibfnamefont {C.}~\bibnamefont {Sun}}, \bibinfo
			{author} {\bibfnamefont {N.}~\bibnamefont {Wang}}, \ and\ \bibinfo {author}
			{\bibfnamefont {Q.}~\bibnamefont {Zhang}},\ }\href@noop {} {\bibfield
			{journal} {\bibinfo  {journal} {J. Mater. Chem. A}\ }\textbf {\bibinfo
				{volume} {7}},\ \bibinfo {pages} {10138} (\bibinfo {year}
			{2019}{\natexlab{a}})}\BibitemShut {NoStop}%
		\bibitem [{\citenamefont {Liu}\ \emph {et~al.}(2023)\citenamefont {Liu},
			\citenamefont {Zhang}, \citenamefont {Ma}, \citenamefont {Zhao},
			\citenamefont {Li},\ and\ \citenamefont {Cui}}]{liu2023challenges}%
		\BibitemOpen
		\bibfield  {author} {\bibinfo {author} {\bibfnamefont {Y.}~\bibnamefont
				{Liu}}, \bibinfo {author} {\bibfnamefont {Y.-H.}\ \bibnamefont {Zhang}},
			\bibinfo {author} {\bibfnamefont {J.}~\bibnamefont {Ma}}, \bibinfo {author}
			{\bibfnamefont {J.}~\bibnamefont {Zhao}}, \bibinfo {author} {\bibfnamefont
				{X.}~\bibnamefont {Li}}, \ and\ \bibinfo {author} {\bibfnamefont
				{G.}~\bibnamefont {Cui}},\ }\href@noop {} {\bibfield  {journal} {\bibinfo
				{journal} {Chem. Mater.}\ }\textbf {\bibinfo {volume} {36}},\ \bibinfo
			{pages} {54} (\bibinfo {year} {2023})}\BibitemShut {NoStop}%
		\bibitem [{\citenamefont {Liu}\ \emph {et~al.}(2021{\natexlab{a}})\citenamefont
			{Liu}, \citenamefont {Li},\ and\ \citenamefont {Xia}}]{juyinlizi}%
		\BibitemOpen
		\bibfield  {author} {\bibinfo {author} {\bibfnamefont {Y.}~\bibnamefont
				{Liu}}, \bibinfo {author} {\bibfnamefont {W.}~\bibnamefont {Li}}, \ and\
			\bibinfo {author} {\bibfnamefont {Y.}~\bibnamefont {Xia}},\ }\href@noop {}
		{\bibfield  {journal} {\bibinfo  {journal} {Electrochem. Energy Rev.}\
			}\textbf {\bibinfo {volume} {4}},\ \bibinfo {pages} {447} (\bibinfo {year}
			{2021}{\natexlab{a}})}\BibitemShut {NoStop}%
		\bibitem [{\citenamefont {Peng}\ \emph {et~al.}(2022)\citenamefont {Peng},
			\citenamefont {Zhang}, \citenamefont {Liu}, \citenamefont {Wang},
			\citenamefont {Chou}, \citenamefont {Liu},\ and\ \citenamefont
			{Dou}}]{prussianblue}%
		\BibitemOpen
		\bibfield  {author} {\bibinfo {author} {\bibfnamefont {J.}~\bibnamefont
				{Peng}}, \bibinfo {author} {\bibfnamefont {W.}~\bibnamefont {Zhang}},
			\bibinfo {author} {\bibfnamefont {Q.}~\bibnamefont {Liu}}, \bibinfo {author}
			{\bibfnamefont {J.}~\bibnamefont {Wang}}, \bibinfo {author} {\bibfnamefont
				{S.}~\bibnamefont {Chou}}, \bibinfo {author} {\bibfnamefont {H.}~\bibnamefont
				{Liu}}, \ and\ \bibinfo {author} {\bibfnamefont {S.}~\bibnamefont {Dou}},\
		}\href@noop {} {\bibfield  {journal} {\bibinfo  {journal} {Adv. Mater.}\
			}\textbf {\bibinfo {volume} {34}},\ \bibinfo {pages} {2108384} (\bibinfo
			{year} {2022})}\BibitemShut {NoStop}%
		\bibitem [{\citenamefont {Liu}\ \emph {et~al.}(2024)\citenamefont {Liu},
			\citenamefont {Kong}, \citenamefont {Wang}, \citenamefont {Li}, \citenamefont
			{Wang}, \citenamefont {Zhu}, \citenamefont {Li}, \citenamefont {Jian},
			\citenamefont {Jia}, \citenamefont {Su}, \citenamefont {Zhang}, \citenamefont
			{Mao}, \citenamefont {Chen}, \citenamefont {Liu}, \citenamefont {Chou},\ and\
			\citenamefont {Xiao}}]{liu2024reviving}%
		\BibitemOpen
		\bibfield  {author} {\bibinfo {author} {\bibfnamefont {H.}~\bibnamefont
				{Liu}}, \bibinfo {author} {\bibfnamefont {L.}~\bibnamefont {Kong}}, \bibinfo
			{author} {\bibfnamefont {H.}~\bibnamefont {Wang}}, \bibinfo {author}
			{\bibfnamefont {J.}~\bibnamefont {Li}}, \bibinfo {author} {\bibfnamefont
				{J.}~\bibnamefont {Wang}}, \bibinfo {author} {\bibfnamefont {Y.}~\bibnamefont
				{Zhu}}, \bibinfo {author} {\bibfnamefont {H.}~\bibnamefont {Li}}, \bibinfo
			{author} {\bibfnamefont {Z.}~\bibnamefont {Jian}}, \bibinfo {author}
			{\bibfnamefont {X.}~\bibnamefont {Jia}}, \bibinfo {author} {\bibfnamefont
				{Y.}~\bibnamefont {Su}}, \bibinfo {author} {\bibfnamefont {S.}~\bibnamefont
				{Zhang}}, \bibinfo {author} {\bibfnamefont {J.}~\bibnamefont {Mao}}, \bibinfo
			{author} {\bibfnamefont {S.}~\bibnamefont {Chen}}, \bibinfo {author}
			{\bibfnamefont {Y.}~\bibnamefont {Liu}}, \bibinfo {author} {\bibfnamefont
				{S.}~\bibnamefont {Chou}}, \ and\ \bibinfo {author} {\bibfnamefont
				{Y.}~\bibnamefont {Xiao}},\ }\href@noop {} {\bibfield  {journal} {\bibinfo
				{journal} {Adv. Mater.}\ }\textbf {\bibinfo {volume} {36}},\ \bibinfo {pages}
			{2407994} (\bibinfo {year} {2024})}\BibitemShut {NoStop}%
		\bibitem [{\citenamefont {Xiao}\ \emph {et~al.}(2025)\citenamefont {Xiao},
			\citenamefont {Sun}, \citenamefont {Chen}, \citenamefont {Wang},
			\citenamefont {Ding}, \citenamefont {Tan}, \citenamefont {Sun}, \citenamefont
			{Zhang}, \citenamefont {Wang}, \citenamefont {Mao},\ and\ \citenamefont
			{Zhu}}]{xiao2025guideline}%
		\BibitemOpen
		\bibfield  {author} {\bibinfo {author} {\bibfnamefont {Y.}~\bibnamefont
				{Xiao}}, \bibinfo {author} {\bibfnamefont {Q.-Q.}\ \bibnamefont {Sun}},
			\bibinfo {author} {\bibfnamefont {D.}~\bibnamefont {Chen}}, \bibinfo {author}
			{\bibfnamefont {J.}~\bibnamefont {Wang}}, \bibinfo {author} {\bibfnamefont
				{J.}~\bibnamefont {Ding}}, \bibinfo {author} {\bibfnamefont {P.}~\bibnamefont
				{Tan}}, \bibinfo {author} {\bibfnamefont {Y.}~\bibnamefont {Sun}}, \bibinfo
			{author} {\bibfnamefont {S.}~\bibnamefont {Zhang}}, \bibinfo {author}
			{\bibfnamefont {P.-F.}\ \bibnamefont {Wang}}, \bibinfo {author}
			{\bibfnamefont {J.}~\bibnamefont {Mao}}, \ and\ \bibinfo {author}
			{\bibfnamefont {Y.-F.}\ \bibnamefont {Zhu}},\ }\href@noop {} {\bibfield
			{journal} {\bibinfo  {journal} {Adv. Mater.}\ }\textbf {\bibinfo {volume}
				{37}},\ \bibinfo {pages} {2504312} (\bibinfo {year} {2025})}\BibitemShut
		{NoStop}%
		\bibitem [{\citenamefont {Zeng}\ \emph {et~al.}(2025)\citenamefont {Zeng},
			\citenamefont {Ni}, \citenamefont {Li}, \citenamefont {Cheng}, \citenamefont
			{Shao}, \citenamefont {Feng}, \citenamefont {Niu}, \citenamefont {Zhao},
			\citenamefont {Li}, \citenamefont {Li}, \citenamefont {Xie}, \citenamefont
			{Lu}, \citenamefont {Yan}, \citenamefont {Zhang},\ and\ \citenamefont
			{Chen}}]{zeng2025crystalline}%
		\BibitemOpen
		\bibfield  {author} {\bibinfo {author} {\bibfnamefont {M.}~\bibnamefont
				{Zeng}}, \bibinfo {author} {\bibfnamefont {Y.}~\bibnamefont {Ni}}, \bibinfo
			{author} {\bibfnamefont {Y.}~\bibnamefont {Li}}, \bibinfo {author}
			{\bibfnamefont {W.}~\bibnamefont {Cheng}}, \bibinfo {author} {\bibfnamefont
				{Z.}~\bibnamefont {Shao}}, \bibinfo {author} {\bibfnamefont {X.}~\bibnamefont
				{Feng}}, \bibinfo {author} {\bibfnamefont {Z.}~\bibnamefont {Niu}}, \bibinfo
			{author} {\bibfnamefont {Q.}~\bibnamefont {Zhao}}, \bibinfo {author}
			{\bibfnamefont {Y.-A.}\ \bibnamefont {Li}}, \bibinfo {author} {\bibfnamefont
				{H.}~\bibnamefont {Li}}, \bibinfo {author} {\bibfnamefont {W.}~\bibnamefont
				{Xie}}, \bibinfo {author} {\bibfnamefont {Y.}~\bibnamefont {Lu}}, \bibinfo
			{author} {\bibfnamefont {Z.}~\bibnamefont {Yan}}, \bibinfo {author}
			{\bibfnamefont {K.}~\bibnamefont {Zhang}}, \ and\ \bibinfo {author}
			{\bibfnamefont {J.}~\bibnamefont {Chen}},\ }\href@noop {} {\bibfield
			{journal} {\bibinfo  {journal} {Angew. Chem. Int. Ed}\ }\textbf {\bibinfo
				{volume} {64}},\ \bibinfo {pages} {e202501134} (\bibinfo {year}
			{2025})}\BibitemShut {NoStop}%
		\bibitem [{\citenamefont {Zuo}\ and\ \citenamefont {Yang}(2022)}]{Mn-NaMnO2}%
		\BibitemOpen
		\bibfield  {author} {\bibinfo {author} {\bibfnamefont {W.}~\bibnamefont
				{Zuo}}\ and\ \bibinfo {author} {\bibfnamefont {Y.}~\bibnamefont {Yang}},\
		}\href@noop {} {\bibfield  {journal} {\bibinfo  {journal} {Acc. Mater. Res.}\
			}\textbf {\bibinfo {volume} {3}},\ \bibinfo {pages} {709} (\bibinfo {year}
			{2022})}\BibitemShut {NoStop}%
		\bibitem [{\citenamefont {Bi}\ \emph {et~al.}(2021)\citenamefont {Bi},
			\citenamefont {Wang}, \citenamefont {Yue}, \citenamefont {Tie},\ and\
			\citenamefont {Niu}}]{bi2021rechargeable}%
		\BibitemOpen
		\bibfield  {author} {\bibinfo {author} {\bibfnamefont {S.}~\bibnamefont
				{Bi}}, \bibinfo {author} {\bibfnamefont {S.}~\bibnamefont {Wang}}, \bibinfo
			{author} {\bibfnamefont {F.}~\bibnamefont {Yue}}, \bibinfo {author}
			{\bibfnamefont {Z.}~\bibnamefont {Tie}}, \ and\ \bibinfo {author}
			{\bibfnamefont {Z.}~\bibnamefont {Niu}},\ }\href@noop {} {\bibfield
			{journal} {\bibinfo  {journal} {Nat. Commun.}\ }\textbf {\bibinfo {volume}
				{12}},\ \bibinfo {pages} {6991} (\bibinfo {year} {2021})}\BibitemShut
		{NoStop}%
		\bibitem [{\citenamefont {Zhang}\ \emph {et~al.}(2025)\citenamefont {Zhang},
			\citenamefont {Yin}, \citenamefont {Ning}, \citenamefont {Chai},
			\citenamefont {Du}, \citenamefont {Hao}, \citenamefont {Wang}, \citenamefont
			{Liu}, \citenamefont {Gao}, \citenamefont {Wang}, , \citenamefont {Yao},
			\citenamefont {Li},\ and\ \citenamefont {Zhou}}]{zhang2025crystal}%
		\BibitemOpen
		\bibfield  {author} {\bibinfo {author} {\bibfnamefont {G.}~\bibnamefont
				{Zhang}}, \bibinfo {author} {\bibfnamefont {X.}~\bibnamefont {Yin}}, \bibinfo
			{author} {\bibfnamefont {D.}~\bibnamefont {Ning}}, \bibinfo {author}
			{\bibfnamefont {Y.}~\bibnamefont {Chai}}, \bibinfo {author} {\bibfnamefont
				{R.}~\bibnamefont {Du}}, \bibinfo {author} {\bibfnamefont {D.}~\bibnamefont
				{Hao}}, \bibinfo {author} {\bibfnamefont {C.}~\bibnamefont {Wang}}, \bibinfo
			{author} {\bibfnamefont {X.}~\bibnamefont {Liu}}, \bibinfo {author}
			{\bibfnamefont {R.}~\bibnamefont {Gao}}, \bibinfo {author} {\bibfnamefont
				{J.}~\bibnamefont {Wang}}, , \bibinfo {author} {\bibfnamefont
				{X.}~\bibnamefont {Yao}}, \bibinfo {author} {\bibfnamefont {Y.}~\bibnamefont
				{Li}}, \ and\ \bibinfo {author} {\bibfnamefont {D.}~\bibnamefont {Zhou}},\
		}\href@noop {} {\bibfield  {journal} {\bibinfo  {journal} {Angew. Chem. Int.
					Ed.}\ }\textbf {\bibinfo {volume} {64}},\ \bibinfo {pages} {e202415450}
			(\bibinfo {year} {2025})}\BibitemShut {NoStop}%
		\bibitem [{\citenamefont {Delmas}\ \emph {et~al.}(1980)\citenamefont {Delmas},
			\citenamefont {Fouassier},\ and\ \citenamefont {Hagenmuller}}]{name}%
		\BibitemOpen
		\bibfield  {author} {\bibinfo {author} {\bibfnamefont {C.}~\bibnamefont
				{Delmas}}, \bibinfo {author} {\bibfnamefont {C.}~\bibnamefont {Fouassier}}, \
			and\ \bibinfo {author} {\bibfnamefont {P.}~\bibnamefont {Hagenmuller}},\
		}\href@noop {} {\bibfield  {journal} {\bibinfo  {journal} {Phys. B+C}\
			}\textbf {\bibinfo {volume} {99}},\ \bibinfo {pages} {81} (\bibinfo {year}
			{1980})}\BibitemShut {NoStop}%
		\bibitem [{\citenamefont {Zuo}\ \emph {et~al.}(2020)\citenamefont {Zuo},
			\citenamefont {Qiu}, \citenamefont {Liu}, \citenamefont {Zheng},
			\citenamefont {Zhao}, \citenamefont {Li}, \citenamefont {He}, \citenamefont
			{Zhou}, \citenamefont {Xiao}, \citenamefont {Li}, \citenamefont {Ortiz},\
			and\ \citenamefont {Yang}}]{P'2}%
		\BibitemOpen
		\bibfield  {author} {\bibinfo {author} {\bibfnamefont {W.}~\bibnamefont
				{Zuo}}, \bibinfo {author} {\bibfnamefont {J.}~\bibnamefont {Qiu}}, \bibinfo
			{author} {\bibfnamefont {X.}~\bibnamefont {Liu}}, \bibinfo {author}
			{\bibfnamefont {B.}~\bibnamefont {Zheng}}, \bibinfo {author} {\bibfnamefont
				{Y.}~\bibnamefont {Zhao}}, \bibinfo {author} {\bibfnamefont {J.}~\bibnamefont
				{Li}}, \bibinfo {author} {\bibfnamefont {H.}~\bibnamefont {He}}, \bibinfo
			{author} {\bibfnamefont {K.}~\bibnamefont {Zhou}}, \bibinfo {author}
			{\bibfnamefont {Z.}~\bibnamefont {Xiao}}, \bibinfo {author} {\bibfnamefont
				{Q.}~\bibnamefont {Li}}, \bibinfo {author} {\bibfnamefont {G.~F.}\
				\bibnamefont {Ortiz}}, \ and\ \bibinfo {author} {\bibfnamefont
				{Y.}~\bibnamefont {Yang}},\ }\href@noop {} {\bibfield  {journal} {\bibinfo
				{journal} {Energy Storage Mater.}\ }\textbf {\bibinfo {volume} {26}},\
			\bibinfo {pages} {503} (\bibinfo {year} {2020})}\BibitemShut {NoStop}%
		\bibitem [{\citenamefont {Luong}\ \emph {et~al.}(2020)\citenamefont {Luong},
			\citenamefont {Dinh}, \citenamefont {Momida},\ and\ \citenamefont
			{Oguchi}}]{Cmcm-DFT-pccp-2020}%
		\BibitemOpen
		\bibfield  {author} {\bibinfo {author} {\bibfnamefont {H.~D.}\ \bibnamefont
				{Luong}}, \bibinfo {author} {\bibfnamefont {V.~A.}\ \bibnamefont {Dinh}},
			\bibinfo {author} {\bibfnamefont {H.}~\bibnamefont {Momida}}, \ and\ \bibinfo
			{author} {\bibfnamefont {T.}~\bibnamefont {Oguchi}},\ }\href@noop {}
		{\bibfield  {journal} {\bibinfo  {journal} {Phys. Chem. Chem. Phys.}\
			}\textbf {\bibinfo {volume} {22}},\ \bibinfo {pages} {18219} (\bibinfo {year}
			{2020})}\BibitemShut {NoStop}%
		\bibitem [{\citenamefont {Wang}\ \emph {et~al.}(2023)\citenamefont {Wang},
			\citenamefont {Fu}, \citenamefont {Liu}, \citenamefont {Saravanan},
			\citenamefont {Luo}, \citenamefont {Deng}, \citenamefont {Sham},
			\citenamefont {Sun},\ and\ \citenamefont {Mo}}]{PvsO}%
		\BibitemOpen
		\bibfield  {author} {\bibinfo {author} {\bibfnamefont {S.}~\bibnamefont
				{Wang}}, \bibinfo {author} {\bibfnamefont {J.}~\bibnamefont {Fu}}, \bibinfo
			{author} {\bibfnamefont {Y.}~\bibnamefont {Liu}}, \bibinfo {author}
			{\bibfnamefont {R.~S.}\ \bibnamefont {Saravanan}}, \bibinfo {author}
			{\bibfnamefont {J.}~\bibnamefont {Luo}}, \bibinfo {author} {\bibfnamefont
				{S.}~\bibnamefont {Deng}}, \bibinfo {author} {\bibfnamefont {T.-K.}\
				\bibnamefont {Sham}}, \bibinfo {author} {\bibfnamefont {X.}~\bibnamefont
				{Sun}}, \ and\ \bibinfo {author} {\bibfnamefont {Y.}~\bibnamefont {Mo}},\
		}\href@noop {} {\bibfield  {journal} {\bibinfo  {journal} {Nat. Commun.}\
			}\textbf {\bibinfo {volume} {14}},\ \bibinfo {pages} {7615} (\bibinfo {year}
			{2023})}\BibitemShut {NoStop}%
		\bibitem [{\citenamefont {Ma}\ \emph {et~al.}(2011)\citenamefont {Ma},
			\citenamefont {Chen},\ and\ \citenamefont {Ceder}}]{O'3}%
		\BibitemOpen
		\bibfield  {author} {\bibinfo {author} {\bibfnamefont {X.}~\bibnamefont
				{Ma}}, \bibinfo {author} {\bibfnamefont {H.}~\bibnamefont {Chen}}, \ and\
			\bibinfo {author} {\bibfnamefont {G.}~\bibnamefont {Ceder}},\ }\href@noop {}
		{\bibfield  {journal} {\bibinfo  {journal} {J. Electrochem. Soc.}\ }\textbf
			{\bibinfo {volume} {158}},\ \bibinfo {pages} {A1307} (\bibinfo {year}
			{2011})}\BibitemShut {NoStop}%
		\bibitem [{\citenamefont {Billaud}\ \emph {et~al.}(2014)\citenamefont
			{Billaud}, \citenamefont {Cl{\'e}ment}, \citenamefont {Armstrong},
			\citenamefont {Canales-V{\'a}zquez}, \citenamefont {Rozier}, \citenamefont
			{Grey},\ and\ \citenamefont {Bruce}}]{beita}%
		\BibitemOpen
		\bibfield  {author} {\bibinfo {author} {\bibfnamefont {J.}~\bibnamefont
				{Billaud}}, \bibinfo {author} {\bibfnamefont {R.~J.}\ \bibnamefont
				{Cl{\'e}ment}}, \bibinfo {author} {\bibfnamefont {A.~R.}\ \bibnamefont
				{Armstrong}}, \bibinfo {author} {\bibfnamefont {J.}~\bibnamefont
				{Canales-V{\'a}zquez}}, \bibinfo {author} {\bibfnamefont {P.}~\bibnamefont
				{Rozier}}, \bibinfo {author} {\bibfnamefont {C.~P.}\ \bibnamefont {Grey}}, \
			and\ \bibinfo {author} {\bibfnamefont {P.~G.}\ \bibnamefont {Bruce}},\
		}\href@noop {} {\bibfield  {journal} {\bibinfo  {journal} {J. Am. Chem.
					Soc.}\ }\textbf {\bibinfo {volume} {136}},\ \bibinfo {pages} {17243}
			(\bibinfo {year} {2014})}\BibitemShut {NoStop}%
		\bibitem [{\citenamefont {Daubner}\ \emph {et~al.}(2024)\citenamefont
			{Daubner}, \citenamefont {Dillenz}, \citenamefont {Pfeiffer}, \citenamefont
			{Gauckler}, \citenamefont {Rosin}, \citenamefont {Burgard}, \citenamefont
			{Martin}, \citenamefont {Axmann}, \citenamefont {Sotoudeh}, \citenamefont
			{Gro{\ss}}, \citenamefont {Schneider},\ and\ \citenamefont
			{Nestler}}]{npj-2024}%
		\BibitemOpen
		\bibfield  {author} {\bibinfo {author} {\bibfnamefont {S.}~\bibnamefont
				{Daubner}}, \bibinfo {author} {\bibfnamefont {M.}~\bibnamefont {Dillenz}},
			\bibinfo {author} {\bibfnamefont {L.~F.}\ \bibnamefont {Pfeiffer}}, \bibinfo
			{author} {\bibfnamefont {C.}~\bibnamefont {Gauckler}}, \bibinfo {author}
			{\bibfnamefont {M.}~\bibnamefont {Rosin}}, \bibinfo {author} {\bibfnamefont
				{N.}~\bibnamefont {Burgard}}, \bibinfo {author} {\bibfnamefont
				{J.}~\bibnamefont {Martin}}, \bibinfo {author} {\bibfnamefont
				{P.}~\bibnamefont {Axmann}}, \bibinfo {author} {\bibfnamefont
				{M.}~\bibnamefont {Sotoudeh}}, \bibinfo {author} {\bibfnamefont
				{A.}~\bibnamefont {Gro{\ss}}}, \bibinfo {author} {\bibfnamefont
				{D.}~\bibnamefont {Schneider}}, \ and\ \bibinfo {author} {\bibfnamefont
				{B.}~\bibnamefont {Nestler}},\ }\href@noop {} {\bibfield  {journal} {\bibinfo
				{journal} {npj Comput. Mater.}\ }\textbf {\bibinfo {volume} {10}},\ \bibinfo
			{pages} {75} (\bibinfo {year} {2024})}\BibitemShut {NoStop}%
		\bibitem [{\citenamefont {Liu}\ \emph {et~al.}(2020)\citenamefont {Liu},
			\citenamefont {Hu}, \citenamefont {Chen}, \citenamefont {Zou}, \citenamefont
			{Jin}, \citenamefont {Wang}, \citenamefont {Chou}, \citenamefont {Liu},\ and\
			\citenamefont {Dou}}]{chanza-de-quedian}%
		\BibitemOpen
		\bibfield  {author} {\bibinfo {author} {\bibfnamefont {Q.}~\bibnamefont
				{Liu}}, \bibinfo {author} {\bibfnamefont {Z.}~\bibnamefont {Hu}}, \bibinfo
			{author} {\bibfnamefont {M.}~\bibnamefont {Chen}}, \bibinfo {author}
			{\bibfnamefont {C.}~\bibnamefont {Zou}}, \bibinfo {author} {\bibfnamefont
				{H.}~\bibnamefont {Jin}}, \bibinfo {author} {\bibfnamefont {S.}~\bibnamefont
				{Wang}}, \bibinfo {author} {\bibfnamefont {S.-L.}\ \bibnamefont {Chou}},
			\bibinfo {author} {\bibfnamefont {Y.}~\bibnamefont {Liu}}, \ and\ \bibinfo
			{author} {\bibfnamefont {S.-X.}\ \bibnamefont {Dou}},\ }\href@noop {}
		{\bibfield  {journal} {\bibinfo  {journal} {Adv. Funct. Mater.}\ }\textbf
			{\bibinfo {volume} {30}},\ \bibinfo {pages} {1909530} (\bibinfo {year}
			{2020})}\BibitemShut {NoStop}%
		\bibitem [{\citenamefont {Liu}\ \emph {et~al.}(2021{\natexlab{b}})\citenamefont
			{Liu}, \citenamefont {Wang}, \citenamefont {Zhang}, \citenamefont {Zhao},
			\citenamefont {Zhang},\ and\ \citenamefont {Yu}}]{Matter-2021}%
		\BibitemOpen
		\bibfield  {author} {\bibinfo {author} {\bibfnamefont {S.}~\bibnamefont
				{Liu}}, \bibinfo {author} {\bibfnamefont {B.}~\bibnamefont {Wang}}, \bibinfo
			{author} {\bibfnamefont {X.}~\bibnamefont {Zhang}}, \bibinfo {author}
			{\bibfnamefont {S.}~\bibnamefont {Zhao}}, \bibinfo {author} {\bibfnamefont
				{Z.}~\bibnamefont {Zhang}}, \ and\ \bibinfo {author} {\bibfnamefont
				{H.}~\bibnamefont {Yu}},\ }\href@noop {} {\bibfield  {journal} {\bibinfo
				{journal} {Matter}\ }\textbf {\bibinfo {volume} {4}},\ \bibinfo {pages}
			{1511} (\bibinfo {year} {2021}{\natexlab{b}})}\BibitemShut {NoStop}%
		\bibitem [{\citenamefont {Uyama}\ \emph {et~al.}(2018)\citenamefont {Uyama},
			\citenamefont {Mukai},\ and\ \citenamefont {Yamada}}]{LiMnO2}%
		\BibitemOpen
		\bibfield  {author} {\bibinfo {author} {\bibfnamefont {T.}~\bibnamefont
				{Uyama}}, \bibinfo {author} {\bibfnamefont {K.}~\bibnamefont {Mukai}}, \ and\
			\bibinfo {author} {\bibfnamefont {I.}~\bibnamefont {Yamada}},\ }\href@noop {}
		{\bibfield  {journal} {\bibinfo  {journal} {RSC Adv.}\ }\textbf {\bibinfo
				{volume} {8}},\ \bibinfo {pages} {26325} (\bibinfo {year}
			{2018})}\BibitemShut {NoStop}%
		\bibitem [{\citenamefont {Wang}\ \emph
			{et~al.}(2019{\natexlab{b}})\citenamefont {Wang}, \citenamefont {Liu},
			\citenamefont {Qie}, \citenamefont {Gong}, \citenamefont {Zhang},
			\citenamefont {Sun},\ and\ \citenamefont {Jena}}]{wangshuo-Li2MnSiO4}%
		\BibitemOpen
		\bibfield  {author} {\bibinfo {author} {\bibfnamefont {S.}~\bibnamefont
				{Wang}}, \bibinfo {author} {\bibfnamefont {J.}~\bibnamefont {Liu}}, \bibinfo
			{author} {\bibfnamefont {Y.}~\bibnamefont {Qie}}, \bibinfo {author}
			{\bibfnamefont {S.}~\bibnamefont {Gong}}, \bibinfo {author} {\bibfnamefont
				{C.}~\bibnamefont {Zhang}}, \bibinfo {author} {\bibfnamefont
				{Q.}~\bibnamefont {Sun}}, \ and\ \bibinfo {author} {\bibfnamefont
				{P.}~\bibnamefont {Jena}},\ }\href@noop {} {\bibfield  {journal} {\bibinfo
				{journal} {J. Mater. Chem. A}\ }\textbf {\bibinfo {volume} {7}},\ \bibinfo
			{pages} {16406} (\bibinfo {year} {2019}{\natexlab{b}})}\BibitemShut {NoStop}%
		\bibitem [{\citenamefont {Wang}\ \emph {et~al.}(2018)\citenamefont {Wang},
			\citenamefont {Liu}, \citenamefont {Qie}, \citenamefont {Gong}, \citenamefont
			{Sun},\ and\ \citenamefont {Jena}}]{wangshuo-LiCoO2}%
		\BibitemOpen
		\bibfield  {author} {\bibinfo {author} {\bibfnamefont {S.}~\bibnamefont
				{Wang}}, \bibinfo {author} {\bibfnamefont {J.}~\bibnamefont {Liu}}, \bibinfo
			{author} {\bibfnamefont {Y.}~\bibnamefont {Qie}}, \bibinfo {author}
			{\bibfnamefont {S.}~\bibnamefont {Gong}}, \bibinfo {author} {\bibfnamefont
				{Q.}~\bibnamefont {Sun}}, \ and\ \bibinfo {author} {\bibfnamefont
				{P.}~\bibnamefont {Jena}},\ }\href@noop {} {\bibfield  {journal} {\bibinfo
				{journal} {J. Mater. Chem. A}\ }\textbf {\bibinfo {volume} {6}},\ \bibinfo
			{pages} {18449} (\bibinfo {year} {2018})}\BibitemShut {NoStop}%
		\bibitem [{\citenamefont {Wang}\ \emph {et~al.}(2017)\citenamefont {Wang},
			\citenamefont {Liu},\ and\ \citenamefont {Sun}}]{wangshuo-Li2MnO3}%
		\BibitemOpen
		\bibfield  {author} {\bibinfo {author} {\bibfnamefont {S.}~\bibnamefont
				{Wang}}, \bibinfo {author} {\bibfnamefont {J.}~\bibnamefont {Liu}}, \ and\
			\bibinfo {author} {\bibfnamefont {Q.}~\bibnamefont {Sun}},\ }\href@noop {}
		{\bibfield  {journal} {\bibinfo  {journal} {J. Mater. Chem. A}\ }\textbf
			{\bibinfo {volume} {5}},\ \bibinfo {pages} {16936} (\bibinfo {year}
			{2017})}\BibitemShut {NoStop}%
		\bibitem [{\citenamefont {Lonie}\ and\ \citenamefont
			{Zurek}(2011)}]{lonie2011xtalopt}%
		\BibitemOpen
		\bibfield  {author} {\bibinfo {author} {\bibfnamefont {D.~C.}\ \bibnamefont
				{Lonie}}\ and\ \bibinfo {author} {\bibfnamefont {E.}~\bibnamefont {Zurek}},\
		}\href@noop {} {\bibfield  {journal} {\bibinfo  {journal} {Comput. Phys.
					Commun.}\ }\textbf {\bibinfo {volume} {182}},\ \bibinfo {pages} {372}
			(\bibinfo {year} {2011})}\BibitemShut {NoStop}%
		\bibitem [{\citenamefont {Hajinazar}\ and\ \citenamefont
			{Zurek}(2024)}]{hajinazar2024xtalopt}%
		\BibitemOpen
		\bibfield  {author} {\bibinfo {author} {\bibfnamefont {S.}~\bibnamefont
				{Hajinazar}}\ and\ \bibinfo {author} {\bibfnamefont {E.}~\bibnamefont
				{Zurek}},\ }\href@noop {} {\bibfield  {journal} {\bibinfo  {journal} {Comput.
					Phys. Commun.}\ }\textbf {\bibinfo {volume} {304}},\ \bibinfo {pages}
			{109306} (\bibinfo {year} {2024})}\BibitemShut {NoStop}%
		\bibitem [{\citenamefont {Kresse}\ and\ \citenamefont
			{Furthm{\"u}ller}(1996{\natexlab{a}})}]{VASP}%
		\BibitemOpen
		\bibfield  {author} {\bibinfo {author} {\bibfnamefont {G.}~\bibnamefont
				{Kresse}}\ and\ \bibinfo {author} {\bibfnamefont {J.}~\bibnamefont
				{Furthm{\"u}ller}},\ }\href@noop {} {\bibfield  {journal} {\bibinfo
				{journal} {Phys. Rev. B}\ }\textbf {\bibinfo {volume} {54}},\ \bibinfo
			{pages} {11169} (\bibinfo {year} {1996}{\natexlab{a}})}\BibitemShut {NoStop}%
		\bibitem [{\citenamefont {Kresse}\ and\ \citenamefont
			{Furthm{\"u}ller}(1996{\natexlab{b}})}]{VASP2}%
		\BibitemOpen
		\bibfield  {author} {\bibinfo {author} {\bibfnamefont {G.}~\bibnamefont
				{Kresse}}\ and\ \bibinfo {author} {\bibfnamefont {J.}~\bibnamefont
				{Furthm{\"u}ller}},\ }\href@noop {} {\bibfield  {journal} {\bibinfo
				{journal} {Comput. Mater. Sci}\ }\textbf {\bibinfo {volume} {6}},\ \bibinfo
			{pages} {15} (\bibinfo {year} {1996}{\natexlab{b}})}\BibitemShut {NoStop}%
		\bibitem [{\citenamefont {Bl{\"o}chl}(1994)}]{PBE1}%
		\BibitemOpen
		\bibfield  {author} {\bibinfo {author} {\bibfnamefont {P.~E.}\ \bibnamefont
				{Bl{\"o}chl}},\ }\href@noop {} {\bibfield  {journal} {\bibinfo  {journal}
				{Phys. Rev. B}\ }\textbf {\bibinfo {volume} {50}},\ \bibinfo {pages} {17953}
			(\bibinfo {year} {1994})}\BibitemShut {NoStop}%
		\bibitem [{\citenamefont {Kresse}\ and\ \citenamefont {Joubert}(1999)}]{PBE2}%
		\BibitemOpen
		\bibfield  {author} {\bibinfo {author} {\bibfnamefont {G.}~\bibnamefont
				{Kresse}}\ and\ \bibinfo {author} {\bibfnamefont {D.}~\bibnamefont
				{Joubert}},\ }\href@noop {} {\bibfield  {journal} {\bibinfo  {journal} {Phys.
					Rev. B}\ }\textbf {\bibinfo {volume} {59}},\ \bibinfo {pages} {1758}
			(\bibinfo {year} {1999})}\BibitemShut {NoStop}%
		\bibitem [{\citenamefont {Perdew}\ \emph {et~al.}(1996)\citenamefont {Perdew},
			\citenamefont {Burke},\ and\ \citenamefont {Ernzerhof}}]{PAW}%
		\BibitemOpen
		\bibfield  {author} {\bibinfo {author} {\bibfnamefont {J.~P.}\ \bibnamefont
				{Perdew}}, \bibinfo {author} {\bibfnamefont {K.}~\bibnamefont {Burke}}, \
			and\ \bibinfo {author} {\bibfnamefont {M.}~\bibnamefont {Ernzerhof}},\
		}\href@noop {} {\bibfield  {journal} {\bibinfo  {journal} {Phys. Rev. Lett.}\
			}\textbf {\bibinfo {volume} {77}},\ \bibinfo {pages} {3865} (\bibinfo {year}
			{1996})}\BibitemShut {NoStop}%
		\bibitem [{\citenamefont {Monkhorst}\ and\ \citenamefont
			{Pack}(1976)}]{KPOINTS}%
		\BibitemOpen
		\bibfield  {author} {\bibinfo {author} {\bibfnamefont {H.~J.}\ \bibnamefont
				{Monkhorst}}\ and\ \bibinfo {author} {\bibfnamefont {J.~D.}\ \bibnamefont
				{Pack}},\ }\href@noop {} {\bibfield  {journal} {\bibinfo  {journal} {Phys.
					Rev. B}\ }\textbf {\bibinfo {volume} {13}},\ \bibinfo {pages} {5188}
			(\bibinfo {year} {1976})}\BibitemShut {NoStop}%
		\bibitem [{\citenamefont {Zhu}\ \emph {et~al.}(2019)\citenamefont {Zhu},
			\citenamefont {Peelaers},\ and\ \citenamefont {Van~de
				Walle}}]{U1-cengzhuangnengleicankao}%
		\BibitemOpen
		\bibfield  {author} {\bibinfo {author} {\bibfnamefont {Z.}~\bibnamefont
				{Zhu}}, \bibinfo {author} {\bibfnamefont {H.}~\bibnamefont {Peelaers}}, \
			and\ \bibinfo {author} {\bibfnamefont {C.~G.}\ \bibnamefont {Van~de Walle}},\
		}\href@noop {} {\bibfield  {journal} {\bibinfo  {journal} {Chem. Mater.}\
			}\textbf {\bibinfo {volume} {31}},\ \bibinfo {pages} {5224} (\bibinfo {year}
			{2019})}\BibitemShut {NoStop}%
		\bibitem [{\citenamefont {Cl{\'e}ment}\ \emph {et~al.}(2016)\citenamefont
			{Cl{\'e}ment}, \citenamefont {Middlemiss}, \citenamefont {Seymour},
			\citenamefont {Ilott},\ and\ \citenamefont {Grey}}]{U2-compare}%
		\BibitemOpen
		\bibfield  {author} {\bibinfo {author} {\bibfnamefont {R.~J.}\ \bibnamefont
				{Cl{\'e}ment}}, \bibinfo {author} {\bibfnamefont {D.~S.}\ \bibnamefont
				{Middlemiss}}, \bibinfo {author} {\bibfnamefont {I.~D.}\ \bibnamefont
				{Seymour}}, \bibinfo {author} {\bibfnamefont {A.~J.}\ \bibnamefont {Ilott}},
			\ and\ \bibinfo {author} {\bibfnamefont {C.~P.}\ \bibnamefont {Grey}},\
		}\href@noop {} {\bibfield  {journal} {\bibinfo  {journal} {Chem. Mater.}\
			}\textbf {\bibinfo {volume} {28}},\ \bibinfo {pages} {8228} (\bibinfo {year}
			{2016})}\BibitemShut {NoStop}%
		\bibitem [{\citenamefont {Li}\ \emph {et~al.}(2014)\citenamefont {Li},
			\citenamefont {Ma}, \citenamefont {Su}, \citenamefont {Liu}, \citenamefont
			{Chisnell}, \citenamefont {Ong}, \citenamefont {Chen}, \citenamefont
			{Toumar}, \citenamefont {Idrobo}, \citenamefont {Lei}, \citenamefont {Bai},
			\citenamefont {Wang}, \citenamefont {Lynn}, \citenamefont {Lee},\ and\
			\citenamefont {Ceder}}]{U3}%
		\BibitemOpen
		\bibfield  {author} {\bibinfo {author} {\bibfnamefont {X.}~\bibnamefont
				{Li}}, \bibinfo {author} {\bibfnamefont {X.}~\bibnamefont {Ma}}, \bibinfo
			{author} {\bibfnamefont {D.}~\bibnamefont {Su}}, \bibinfo {author}
			{\bibfnamefont {L.}~\bibnamefont {Liu}}, \bibinfo {author} {\bibfnamefont
				{R.}~\bibnamefont {Chisnell}}, \bibinfo {author} {\bibfnamefont {S.~P.}\
				\bibnamefont {Ong}}, \bibinfo {author} {\bibfnamefont {H.}~\bibnamefont
				{Chen}}, \bibinfo {author} {\bibfnamefont {A.}~\bibnamefont {Toumar}},
			\bibinfo {author} {\bibfnamefont {J.-C.}\ \bibnamefont {Idrobo}}, \bibinfo
			{author} {\bibfnamefont {Y.}~\bibnamefont {Lei}}, \bibinfo {author}
			{\bibfnamefont {J.}~\bibnamefont {Bai}}, \bibinfo {author} {\bibfnamefont
				{F.}~\bibnamefont {Wang}}, \bibinfo {author} {\bibfnamefont {J.~W.}\
				\bibnamefont {Lynn}}, \bibinfo {author} {\bibfnamefont {Y.~S.}\ \bibnamefont
				{Lee}}, \ and\ \bibinfo {author} {\bibfnamefont {G.}~\bibnamefont {Ceder}},\
		}\href@noop {} {\bibfield  {journal} {\bibinfo  {journal} {Nat. Mater.}\
			}\textbf {\bibinfo {volume} {13}},\ \bibinfo {pages} {586} (\bibinfo {year}
			{2014})}\BibitemShut {NoStop}%
		\bibitem [{\citenamefont {Singh}(1997)}]{AFMcankao1}%
		\BibitemOpen
		\bibfield  {author} {\bibinfo {author} {\bibfnamefont {D.~J.}\ \bibnamefont
				{Singh}},\ }\href@noop {} {\bibfield  {journal} {\bibinfo  {journal} {Phys.
					Rev. B}\ }\textbf {\bibinfo {volume} {55}},\ \bibinfo {pages} {309} (\bibinfo
			{year} {1997})}\BibitemShut {NoStop}%
		\bibitem [{\citenamefont {Kam}\ \emph {et~al.}(2025)\citenamefont {Kam},
			\citenamefont {Binci}, \citenamefont {Kaplan}, \citenamefont {Persson},
			\citenamefont {Marzari},\ and\ \citenamefont {Ceder}}]{Ceder-2025-LiMnO2}%
		\BibitemOpen
		\bibfield  {author} {\bibinfo {author} {\bibfnamefont {R.~L.}\ \bibnamefont
				{Kam}}, \bibinfo {author} {\bibfnamefont {L.}~\bibnamefont {Binci}}, \bibinfo
			{author} {\bibfnamefont {A.~D.}\ \bibnamefont {Kaplan}}, \bibinfo {author}
			{\bibfnamefont {K.~A.}\ \bibnamefont {Persson}}, \bibinfo {author}
			{\bibfnamefont {N.}~\bibnamefont {Marzari}}, \ and\ \bibinfo {author}
			{\bibfnamefont {G.}~\bibnamefont {Ceder}},\ }\href@noop {} {\bibfield
			{journal} {\bibinfo  {journal} {Phys. Rev. B}\ }\textbf {\bibinfo {volume}
				{111}},\ \bibinfo {pages} {245132} (\bibinfo {year} {2025})}\BibitemShut
		{NoStop}%
		\bibitem [{\citenamefont {Togo}\ \emph {et~al.}(2008)\citenamefont {Togo},
			\citenamefont {Oba},\ and\ \citenamefont {Tanaka}}]{Phonopy}%
		\BibitemOpen
		\bibfield  {author} {\bibinfo {author} {\bibfnamefont {A.}~\bibnamefont
				{Togo}}, \bibinfo {author} {\bibfnamefont {F.}~\bibnamefont {Oba}}, \ and\
			\bibinfo {author} {\bibfnamefont {I.}~\bibnamefont {Tanaka}},\ }\href@noop {}
		{\bibfield  {journal} {\bibinfo  {journal} {Phys. Rev. B}\ }\textbf {\bibinfo
				{volume} {78}},\ \bibinfo {pages} {134106} (\bibinfo {year}
			{2008})}\BibitemShut {NoStop}%
		\bibitem [{\citenamefont {Nos{\'e}}(1984)}]{Nose-Hoover}%
		\BibitemOpen
		\bibfield  {author} {\bibinfo {author} {\bibfnamefont {S.}~\bibnamefont
				{Nos{\'e}}},\ }\href@noop {} {\bibfield  {journal} {\bibinfo  {journal} {J.
					Chem. Phys.}\ }\textbf {\bibinfo {volume} {81}},\ \bibinfo {pages} {511}
			(\bibinfo {year} {1984})}\BibitemShut {NoStop}%
		\bibitem [{\citenamefont {Grimme}\ \emph {et~al.}(2011)\citenamefont {Grimme},
			\citenamefont {Ehrlich},\ and\ \citenamefont {Goerigk}}]{VDW-D3}%
		\BibitemOpen
		\bibfield  {author} {\bibinfo {author} {\bibfnamefont {S.}~\bibnamefont
				{Grimme}}, \bibinfo {author} {\bibfnamefont {S.}~\bibnamefont {Ehrlich}}, \
			and\ \bibinfo {author} {\bibfnamefont {L.}~\bibnamefont {Goerigk}},\
		}\href@noop {} {\bibfield  {journal} {\bibinfo  {journal} {J. Comput. Chem.}\
			}\textbf {\bibinfo {volume} {32}},\ \bibinfo {pages} {1456} (\bibinfo {year}
			{2011})}\BibitemShut {NoStop}%
		\bibitem [{\citenamefont {Henkelman}\ \emph {et~al.}(2000)\citenamefont
			{Henkelman}, \citenamefont {Uberuaga},\ and\ \citenamefont
			{J{\'o}nsson}}]{CI-NEB}%
		\BibitemOpen
		\bibfield  {author} {\bibinfo {author} {\bibfnamefont {G.}~\bibnamefont
				{Henkelman}}, \bibinfo {author} {\bibfnamefont {B.~P.}\ \bibnamefont
				{Uberuaga}}, \ and\ \bibinfo {author} {\bibfnamefont {H.}~\bibnamefont
				{J{\'o}nsson}},\ }\href@noop {} {\bibfield  {journal} {\bibinfo  {journal}
				{J. Chem. Phys.}\ }\textbf {\bibinfo {volume} {113}},\ \bibinfo {pages}
			{9901} (\bibinfo {year} {2000})}\BibitemShut {NoStop}%
		\bibitem [{\citenamefont {Urban}\ \emph {et~al.}(2016)\citenamefont {Urban},
			\citenamefont {Seo},\ and\ \citenamefont {Ceder}}]{npj-ceder-2016}%
		\BibitemOpen
		\bibfield  {author} {\bibinfo {author} {\bibfnamefont {A.}~\bibnamefont
				{Urban}}, \bibinfo {author} {\bibfnamefont {D.-H.}\ \bibnamefont {Seo}}, \
			and\ \bibinfo {author} {\bibfnamefont {G.}~\bibnamefont {Ceder}},\
		}\href@noop {} {\bibfield  {journal} {\bibinfo  {journal} {npj Comput.
					Mater.}\ }\textbf {\bibinfo {volume} {2}},\ \bibinfo {pages} {1} (\bibinfo
			{year} {2016})}\BibitemShut {NoStop}%
		\bibitem [{\citenamefont {Okhotnikov}\ \emph {et~al.}(2016)\citenamefont
			{Okhotnikov}, \citenamefont {Charpentier},\ and\ \citenamefont
			{Cadars}}]{supercell-program}%
		\BibitemOpen
		\bibfield  {author} {\bibinfo {author} {\bibfnamefont {K.}~\bibnamefont
				{Okhotnikov}}, \bibinfo {author} {\bibfnamefont {T.}~\bibnamefont
				{Charpentier}}, \ and\ \bibinfo {author} {\bibfnamefont {S.}~\bibnamefont
				{Cadars}},\ }\href@noop {} {\bibfield  {journal} {\bibinfo  {journal} {J.
					Cheminf.}\ }\textbf {\bibinfo {volume} {8}},\ \bibinfo {pages} {1} (\bibinfo
			{year} {2016})}\BibitemShut {NoStop}%
		\bibitem [{\citenamefont {Velikokhatnyi}\ \emph {et~al.}(2003)\citenamefont
			{Velikokhatnyi}, \citenamefont {Chang},\ and\ \citenamefont
			{Kumta}}]{AFMcankao3-compare-daixicankao}%
		\BibitemOpen
		\bibfield  {author} {\bibinfo {author} {\bibfnamefont {O.}~\bibnamefont
				{Velikokhatnyi}}, \bibinfo {author} {\bibfnamefont {C.-C.}\ \bibnamefont
				{Chang}}, \ and\ \bibinfo {author} {\bibfnamefont {P.}~\bibnamefont
				{Kumta}},\ }\href@noop {} {\bibfield  {journal} {\bibinfo  {journal} {J.
					Electrochem. Soc.}\ }\textbf {\bibinfo {volume} {150}},\ \bibinfo {pages}
			{A1262} (\bibinfo {year} {2003})}\BibitemShut {NoStop}%
		\bibitem [{\citenamefont {Huang}\ \emph {et~al.}(2014)\citenamefont {Huang},
			\citenamefont {Yu}, \citenamefont {Xu}, \citenamefont {Hu}, \citenamefont
			{Ma}, \citenamefont {Wang}, \citenamefont {Zhao}, \citenamefont {Wen},
			\citenamefont {He}, \citenamefont {Liu},\ and\ \citenamefont
			{Tian}}]{High-pressure-shiyan}%
		\BibitemOpen
		\bibfield  {author} {\bibinfo {author} {\bibfnamefont {Q.}~\bibnamefont
				{Huang}}, \bibinfo {author} {\bibfnamefont {D.}~\bibnamefont {Yu}}, \bibinfo
			{author} {\bibfnamefont {B.}~\bibnamefont {Xu}}, \bibinfo {author}
			{\bibfnamefont {W.}~\bibnamefont {Hu}}, \bibinfo {author} {\bibfnamefont
				{Y.}~\bibnamefont {Ma}}, \bibinfo {author} {\bibfnamefont {Y.}~\bibnamefont
				{Wang}}, \bibinfo {author} {\bibfnamefont {Z.}~\bibnamefont {Zhao}}, \bibinfo
			{author} {\bibfnamefont {B.}~\bibnamefont {Wen}}, \bibinfo {author}
			{\bibfnamefont {J.}~\bibnamefont {He}}, \bibinfo {author} {\bibfnamefont
				{Z.}~\bibnamefont {Liu}}, \ and\ \bibinfo {author} {\bibfnamefont
				{Y.}~\bibnamefont {Tian}},\ }\href@noop {} {\bibfield  {journal} {\bibinfo
				{journal} {Nature}\ }\textbf {\bibinfo {volume} {510}},\ \bibinfo {pages}
			{250} (\bibinfo {year} {2014})}\BibitemShut {NoStop}%
		\bibitem [{\citenamefont {Shishkin}\ and\ \citenamefont
			{Sato}(2021)}]{AFMcankao2}%
		\BibitemOpen
		\bibfield  {author} {\bibinfo {author} {\bibfnamefont {M.}~\bibnamefont
				{Shishkin}}\ and\ \bibinfo {author} {\bibfnamefont {H.}~\bibnamefont
				{Sato}},\ }\href@noop {} {\bibfield  {journal} {\bibinfo  {journal} {J. Phys.
					Chem. C}\ }\textbf {\bibinfo {volume} {125}},\ \bibinfo {pages} {1531}
			(\bibinfo {year} {2021})}\BibitemShut {NoStop}%
		\bibitem [{\citenamefont {Wadhwa}\ \emph {et~al.}(2025)\citenamefont {Wadhwa},
			\citenamefont {Cappellini}, \citenamefont {Teles},\ and\ \citenamefont
			{Filippetti}}]{bandgap-JMCC-2025}%
		\BibitemOpen
		\bibfield  {author} {\bibinfo {author} {\bibfnamefont {P.}~\bibnamefont
				{Wadhwa}}, \bibinfo {author} {\bibfnamefont {G.}~\bibnamefont {Cappellini}},
			\bibinfo {author} {\bibfnamefont {L.}~\bibnamefont {Teles}}, \ and\ \bibinfo
			{author} {\bibfnamefont {A.}~\bibnamefont {Filippetti}},\ }\href@noop {}
		{\bibfield  {journal} {\bibinfo  {journal} {J. Mater. Chem. C}\ }\textbf
			{\bibinfo {volume} {13}},\ \bibinfo {pages} {12483} (\bibinfo {year}
			{2025})}\BibitemShut {NoStop}%
		\bibitem [{\citenamefont {Tian}(1997)}]{prb1997ferrimagnetism}%
		\BibitemOpen
		\bibfield  {author} {\bibinfo {author} {\bibfnamefont {G.-S.}\ \bibnamefont
				{Tian}},\ }\href@noop {} {\bibfield  {journal} {\bibinfo  {journal} {Phys.
					Rev. B}\ }\textbf {\bibinfo {volume} {56}},\ \bibinfo {pages} {5355}
			(\bibinfo {year} {1997})}\BibitemShut {NoStop}%
		\bibitem [{\citenamefont {Kim}\ \emph {et~al.}(2022{\natexlab{a}})\citenamefont
			{Kim}, \citenamefont {Beach}, \citenamefont {Lee}, \citenamefont {Ono},
			\citenamefont {Rasing},\ and\ \citenamefont
			{Yang}}]{nature-materials-Ferrimagnetic-spintronics-2021}%
		\BibitemOpen
		\bibfield  {author} {\bibinfo {author} {\bibfnamefont {S.~K.}\ \bibnamefont
				{Kim}}, \bibinfo {author} {\bibfnamefont {G.~S.}\ \bibnamefont {Beach}},
			\bibinfo {author} {\bibfnamefont {K.-J.}\ \bibnamefont {Lee}}, \bibinfo
			{author} {\bibfnamefont {T.}~\bibnamefont {Ono}}, \bibinfo {author}
			{\bibfnamefont {T.}~\bibnamefont {Rasing}}, \ and\ \bibinfo {author}
			{\bibfnamefont {H.}~\bibnamefont {Yang}},\ }\href@noop {} {\bibfield
			{journal} {\bibinfo  {journal} {Nat. Mater.}\ }\textbf {\bibinfo {volume}
				{21}},\ \bibinfo {pages} {24} (\bibinfo {year}
			{2022}{\natexlab{a}})}\BibitemShut {NoStop}%
		\bibitem [{\citenamefont {Xiao}\ \emph {et~al.}(2012)\citenamefont {Xiao},
			\citenamefont {Li},\ and\ \citenamefont {Chen}}]{Li2MnO3-2012}%
		\BibitemOpen
		\bibfield  {author} {\bibinfo {author} {\bibfnamefont {R.}~\bibnamefont
				{Xiao}}, \bibinfo {author} {\bibfnamefont {H.}~\bibnamefont {Li}}, \ and\
			\bibinfo {author} {\bibfnamefont {L.}~\bibnamefont {Chen}},\ }\href@noop {}
		{\bibfield  {journal} {\bibinfo  {journal} {Chem. Mater.}\ }\textbf {\bibinfo
				{volume} {24}},\ \bibinfo {pages} {4242} (\bibinfo {year}
			{2012})}\BibitemShut {NoStop}%
		\bibitem [{\citenamefont {Parant}\ \emph {et~al.}(1971)\citenamefont {Parant},
			\citenamefont {Olazcuaga}, \citenamefont {Devalette}, \citenamefont
			{Fouassier},\ and\ \citenamefont {Hagenmuller}}]{parant1971quelques}%
		\BibitemOpen
		\bibfield  {author} {\bibinfo {author} {\bibfnamefont {J.-P.}\ \bibnamefont
				{Parant}}, \bibinfo {author} {\bibfnamefont {R.}~\bibnamefont {Olazcuaga}},
			\bibinfo {author} {\bibfnamefont {M.}~\bibnamefont {Devalette}}, \bibinfo
			{author} {\bibfnamefont {C.}~\bibnamefont {Fouassier}}, \ and\ \bibinfo
			{author} {\bibfnamefont {P.}~\bibnamefont {Hagenmuller}},\ }\href@noop {}
		{\bibfield  {journal} {\bibinfo  {journal} {J. Solid State Chem.}\ }\textbf
			{\bibinfo {volume} {3}},\ \bibinfo {pages} {1} (\bibinfo {year}
			{1971})}\BibitemShut {NoStop}%
		\bibitem [{\citenamefont {Guo}\ \emph {et~al.}(2016)\citenamefont {Guo},
			\citenamefont {Sun}, \citenamefont {Yi}, \citenamefont {Zhu}, \citenamefont
			{Liu}, \citenamefont {Zhu}, \citenamefont {Zhu}, \citenamefont {Chen},
			\citenamefont {Ishida},\ and\ \citenamefont {Zhou}}]{vacancy-effect}%
		\BibitemOpen
		\bibfield  {author} {\bibinfo {author} {\bibfnamefont {S.}~\bibnamefont
				{Guo}}, \bibinfo {author} {\bibfnamefont {Y.}~\bibnamefont {Sun}}, \bibinfo
			{author} {\bibfnamefont {J.}~\bibnamefont {Yi}}, \bibinfo {author}
			{\bibfnamefont {K.}~\bibnamefont {Zhu}}, \bibinfo {author} {\bibfnamefont
				{P.}~\bibnamefont {Liu}}, \bibinfo {author} {\bibfnamefont {Y.}~\bibnamefont
				{Zhu}}, \bibinfo {author} {\bibfnamefont {G.-z.}\ \bibnamefont {Zhu}},
			\bibinfo {author} {\bibfnamefont {M.}~\bibnamefont {Chen}}, \bibinfo {author}
			{\bibfnamefont {M.}~\bibnamefont {Ishida}}, \ and\ \bibinfo {author}
			{\bibfnamefont {H.}~\bibnamefont {Zhou}},\ }\href@noop {} {\bibfield
			{journal} {\bibinfo  {journal} {NPG Asia Mater.}\ }\textbf {\bibinfo {volume}
				{8}},\ \bibinfo {pages} {e266} (\bibinfo {year} {2016})}\BibitemShut
		{NoStop}%
		\bibitem [{\citenamefont {Shannon}(1976)}]{Pauling-rules}%
		\BibitemOpen
		\bibfield  {author} {\bibinfo {author} {\bibfnamefont {R.~D.}\ \bibnamefont
				{Shannon}},\ }\href@noop {} {\bibfield  {journal} {\bibinfo  {journal} {Acta
					Crystallogr., Sect. A: Found. Crystallogr.}\ }\textbf {\bibinfo {volume}
				{32}},\ \bibinfo {pages} {751} (\bibinfo {year} {1976})}\BibitemShut
		{NoStop}%
		\bibitem [{\citenamefont {Raty}\ \emph {et~al.}(2007)\citenamefont {Raty},
			\citenamefont {Schwegler},\ and\ \citenamefont {Bonev}}]{bcc-Na-Nature-2007}%
		\BibitemOpen
		\bibfield  {author} {\bibinfo {author} {\bibfnamefont {J.-Y.}\ \bibnamefont
				{Raty}}, \bibinfo {author} {\bibfnamefont {E.}~\bibnamefont {Schwegler}}, \
			and\ \bibinfo {author} {\bibfnamefont {S.~A.}\ \bibnamefont {Bonev}},\
		}\href@noop {} {\bibfield  {journal} {\bibinfo  {journal} {Nature}\ }\textbf
			{\bibinfo {volume} {449}},\ \bibinfo {pages} {448} (\bibinfo {year}
			{2007})}\BibitemShut {NoStop}%
		\bibitem [{\citenamefont {Elatresh}\ \emph {et~al.}(2020)\citenamefont
			{Elatresh}, \citenamefont {Hossain}, \citenamefont {Bhowmick}, \citenamefont
			{Grockowiak}, \citenamefont {Cai}, \citenamefont {Coniglio}, \citenamefont
			{Tozer}, \citenamefont {Ashcroft}, \citenamefont {Bonev}, \citenamefont
			{Deemyad},\ and\ \citenamefont {Hoffmann}}]{bcc-Na-prb-2020}%
		\BibitemOpen
		\bibfield  {author} {\bibinfo {author} {\bibfnamefont {S.}~\bibnamefont
				{Elatresh}}, \bibinfo {author} {\bibfnamefont {M.~T.}\ \bibnamefont
				{Hossain}}, \bibinfo {author} {\bibfnamefont {T.}~\bibnamefont {Bhowmick}},
			\bibinfo {author} {\bibfnamefont {A.}~\bibnamefont {Grockowiak}}, \bibinfo
			{author} {\bibfnamefont {W.}~\bibnamefont {Cai}}, \bibinfo {author}
			{\bibfnamefont {W.}~\bibnamefont {Coniglio}}, \bibinfo {author}
			{\bibfnamefont {S.~W.}\ \bibnamefont {Tozer}}, \bibinfo {author}
			{\bibfnamefont {N.}~\bibnamefont {Ashcroft}}, \bibinfo {author}
			{\bibfnamefont {S.}~\bibnamefont {Bonev}}, \bibinfo {author} {\bibfnamefont
				{S.}~\bibnamefont {Deemyad}}, \ and\ \bibinfo {author} {\bibfnamefont
				{R.}~\bibnamefont {Hoffmann}},\ }\href@noop {} {\bibfield  {journal}
			{\bibinfo  {journal} {Phys. Rev. B}\ }\textbf {\bibinfo {volume} {101}},\
			\bibinfo {pages} {220103} (\bibinfo {year} {2020})}\BibitemShut {NoStop}%
		\bibitem [{\citenamefont {Kim}\ \emph {et~al.}(2022{\natexlab{b}})\citenamefont
			{Kim}, \citenamefont {Kim}, \citenamefont {Cho},\ and\ \citenamefont
			{Kim}}]{oxygen-stability-JMCA-2022}%
		\BibitemOpen
		\bibfield  {author} {\bibinfo {author} {\bibfnamefont {M.}~\bibnamefont
				{Kim}}, \bibinfo {author} {\bibfnamefont {H.}~\bibnamefont {Kim}}, \bibinfo
			{author} {\bibfnamefont {M.}~\bibnamefont {Cho}}, \ and\ \bibinfo {author}
			{\bibfnamefont {D.}~\bibnamefont {Kim}},\ }\href@noop {} {\bibfield
			{journal} {\bibinfo  {journal} {J. Mater. Chem. A}\ }\textbf {\bibinfo
				{volume} {10}},\ \bibinfo {pages} {11101} (\bibinfo {year}
			{2022}{\natexlab{b}})}\BibitemShut {NoStop}%
		\bibitem [{\citenamefont {Bytautas}\ and\ \citenamefont
			{Ruedenberg}(2010)}]{DFT-bad-for-O2-1}%
		\BibitemOpen
		\bibfield  {author} {\bibinfo {author} {\bibfnamefont {L.}~\bibnamefont
				{Bytautas}}\ and\ \bibinfo {author} {\bibfnamefont {K.}~\bibnamefont
				{Ruedenberg}},\ }\href@noop {} {\bibfield  {journal} {\bibinfo  {journal} {J.
					Chem. Phys.}\ }\textbf {\bibinfo {volume} {132}},\ \bibinfo {pages} {074109}
			(\bibinfo {year} {2010})}\BibitemShut {NoStop}%
		\bibitem [{\citenamefont {Bytautas}\ \emph {et~al.}(2010)\citenamefont
			{Bytautas}, \citenamefont {Matsunaga},\ and\ \citenamefont
			{Ruedenberg}}]{DFT-bad-for-O2-2}%
		\BibitemOpen
		\bibfield  {author} {\bibinfo {author} {\bibfnamefont {L.}~\bibnamefont
				{Bytautas}}, \bibinfo {author} {\bibfnamefont {N.}~\bibnamefont {Matsunaga}},
			\ and\ \bibinfo {author} {\bibfnamefont {K.}~\bibnamefont {Ruedenberg}},\
		}\href@noop {} {\bibfield  {journal} {\bibinfo  {journal} {J. Chem. Phys.}\
			}\textbf {\bibinfo {volume} {132}},\ \bibinfo {pages} {074307} (\bibinfo
			{year} {2010})}\BibitemShut {NoStop}%
		\bibitem [{\citenamefont {N{\o}rskov}\ \emph {et~al.}(2004)\citenamefont
			{N{\o}rskov}, \citenamefont {Rossmeisl}, \citenamefont {Logadottir},
			\citenamefont {Lindqvist}, \citenamefont {Kitchin}, \citenamefont
			{Bligaard},\ and\ \citenamefont {Jonsson}}]{experimentally-water}%
		\BibitemOpen
		\bibfield  {author} {\bibinfo {author} {\bibfnamefont {J.~K.}\ \bibnamefont
				{N{\o}rskov}}, \bibinfo {author} {\bibfnamefont {J.}~\bibnamefont
				{Rossmeisl}}, \bibinfo {author} {\bibfnamefont {A.}~\bibnamefont
				{Logadottir}}, \bibinfo {author} {\bibfnamefont {L.}~\bibnamefont
				{Lindqvist}}, \bibinfo {author} {\bibfnamefont {J.~R.}\ \bibnamefont
				{Kitchin}}, \bibinfo {author} {\bibfnamefont {T.}~\bibnamefont {Bligaard}}, \
			and\ \bibinfo {author} {\bibfnamefont {H.}~\bibnamefont {Jonsson}},\
		}\href@noop {} {\bibfield  {journal} {\bibinfo  {journal} {J. Phys. Chem. B}\
			}\textbf {\bibinfo {volume} {108}},\ \bibinfo {pages} {17886} (\bibinfo
			{year} {2004})}\BibitemShut {NoStop}%
		\bibitem [{\citenamefont {Hummelsh{\o}j}\ \emph {et~al.}(2010)\citenamefont
			{Hummelsh{\o}j}, \citenamefont {Blomqvist}, \citenamefont {Datta},
			\citenamefont {Vegge}, \citenamefont {Rossmeisl}, \citenamefont {Thygesen},
			\citenamefont {Luntz}, \citenamefont {Jacobsen},\ and\ \citenamefont
			{N{\o}rskov}}]{Equation-JCP-2010}%
		\BibitemOpen
		\bibfield  {author} {\bibinfo {author} {\bibfnamefont {J.~S.}\ \bibnamefont
				{Hummelsh{\o}j}}, \bibinfo {author} {\bibfnamefont {J.}~\bibnamefont
				{Blomqvist}}, \bibinfo {author} {\bibfnamefont {S.}~\bibnamefont {Datta}},
			\bibinfo {author} {\bibfnamefont {T.}~\bibnamefont {Vegge}}, \bibinfo
			{author} {\bibfnamefont {J.}~\bibnamefont {Rossmeisl}}, \bibinfo {author}
			{\bibfnamefont {K.~S.}\ \bibnamefont {Thygesen}}, \bibinfo {author}
			{\bibfnamefont {A.}~\bibnamefont {Luntz}}, \bibinfo {author} {\bibfnamefont
				{K.~W.}\ \bibnamefont {Jacobsen}}, \ and\ \bibinfo {author} {\bibfnamefont
				{J.~K.}\ \bibnamefont {N{\o}rskov}},\ }\href@noop {} {\bibfield  {journal}
			{\bibinfo  {journal} {J. Chem. Phys.}\ }\textbf {\bibinfo {volume} {132}},\
			\bibinfo {pages} {071101} (\bibinfo {year} {2010})}\BibitemShut {NoStop}%
	\end{thebibliography}
	
	%

\end{document}